\documentclass[11pt]{article}
\pdfoutput=1 
\usepackage{jheppub}
\usepackage[utf8]{inputenc}

\usepackage{graphicx}
\usepackage{caption,epigraph}
\usepackage{subcaption}
\usepackage{hyperref}
\usepackage[T1]{fontenc}
\usepackage[export]{adjustbox}
\usepackage[table]{xcolor}

\usepackage{amssymb}
\usepackage{pifont}

\newcommand{\be}{\begin{equation}}
\newcommand{\ee}{\end{equation}}
\newcommand{\bea}{\begin{eqnarray}}
\newcommand{\eea}{\end{eqnarray}}

\DeclareUnicodeCharacter{2212}{-}
\usepackage{multirow, graphicx,amssymb,url,mathrsfs,amsmath}
\usepackage{amsxtra,amstext,latexsym,dsfont,amsfonts}
\usepackage{color,eucal}
\usepackage{float}
\usepackage{colortbl}
\usepackage{multirow}
\usepackage{tikz}
\usetikzlibrary{tikzmark}
\usetikzlibrary{arrows}
\usepackage{tabularx, array}
\newcolumntype{L}[1]{>{\raggedright\arraybackslash}p{#1}}
\newcolumntype{C}[1]{>{\centering\arraybackslash}p{#1}}
\newcolumntype{R}[1]{>{\raggedleft\arraybackslash}p{#1}}
\makeatother
\makeatletter
\newcommand{\myBig}{\bBigg@{1.75}}
\makeatother

\title{\LARGE Entanglement entropy as an order parameter for strongly coupled nodal line semimetals}

\author[a,b,c]{Matteo Baggioli,}
\author[d,e]{Yan Liu,}
\author[a,b,c]{Xin-Meng Wu}
\affiliation[a]{School of Physics and Astronomy, Shanghai Jiao Tong University, Shanghai 200240, China}
\affiliation[b]{Wilczek Quantum Center, School of Physics and Astronomy, Shanghai Jiao Tong University, Shanghai 200240, China}
\affiliation[c]{Shanghai Research Center for Quantum Sciences, Shanghai 201315, China}
\affiliation[d]{Center for Gravitational Physics, Department of Space Science,
Beihang University, Beijing 100191, China}
\affiliation[e]{Peng Huanwu Collaborative Center for Research and Education, Beihang University, Beijing 100191, China}

\emailAdd{b.matteo@sjtu.edu.cn}
\emailAdd{yanliu@buaa.edu.cn}
\emailAdd{xinmeng.wu@sjtu.edu.cn}

\abstract{Topological semimetals are a class of many-body systems exhibiting novel macroscopic quantum phenomena at the interplay between high energy and condensed matter physics. They display a topological quantum phase transition (TQPT) which evades the standard Landau paradigm. In the case of Weyl semimetals, the anomalous Hall effect is a good non-local order parameter for the TQPT, as it is proportional to the separation between the Weyl nodes in momentum space. On the contrary, for nodal line semimetals (NLSM), the quest for an order parameter is still open. 
By taking advantage of a recently proposed holographic model for strongly-coupled NLSM, we explicitly show that entanglement entropy (EE) provides an optimal probe for nodal topology. We propose a generalized $c$-function, constructed from the EE, as an order parameter for the TQPT. Moreover, we find that the derivative of the renormalized EE with respect to the external coupling driving the TQPT diverges at the critical point, signaling the rise of non-local quantum correlations. Finally, we show that these quantum information quantities are able to characterize not only the critical point but also features of the quantum critical region at finite temperature.
}

\begin{document} 

\maketitle
\section{Introduction} 
 Within the Landau paradigm (LP), phases of matter which exhibit different macroscopic properties are defined by their symmetries, and whether or not those are spontaneously broken \cite{Landau:1937obd}. The LP is not only a powerful classification tool based on the definition of a local order parameter (OP) but also an important ingredient for the identification of the low-energy degrees of freedom, and the understanding of the critical dynamics across classical phase transitions \cite{toledano1987landau,hohenberg2015introduction}. Nevertheless, Nature abounds of apparent exceptions to the LP; topological phases of matter \cite{RevModPhys.89.041004} are the most famous example of this sort. On the other hand, quantum phase transitions defy the LP as well, since they cannot be described in terms of a standard (i.e., local) OP \cite{sachdev_2011,PhysRevB.70.144407}.

One of the attempts to rationalize phases of matter and phase transitions beyond the LP is based on the notion of generalised symmetries (see \cite{McGreevy:2022oyu} for a recent review). An alternative approach utilizes quantum information quantities, such as entanglement entropy \cite{Calabrese:2004eu}, and generalized related concepts (e.g., topological entanglement entropy \cite{PhysRevLett.96.110404,Levin:2006zz}), to describe topological order \cite{Jiang2012} and quantum phase transitions \cite{PhysRevLett.97.050404,Osterloh2002,PhysRevA.69.022107,PhysRevLett.93.086402,PhysRevA.66.032110}.

Topological semimetals (TS) have emerged as a promising platform not only to reach a fundamental understanding of topological quantum many-body systems but also thanks to their incredible potential for applications \cite{Burkov2016}. In TS, the low energy electronic excitations are chiral fermions forming point-shaped (Weyl semimetals, or
WSM) or line-shaped (nodal line semimetals, or NLSM) Fermi surfaces characterized by stable topological invariants. TS undergo quantum critical phase transitions to trivally gapped insulating states driven by non-thermal external parameters, such as uniaxial strain or chemical pressure \cite{PhysRevLett.108.175303,PhysRevLett.118.240403,PhysRevX.8.011049,Tarruell2012,doi:10.1126/science.aaa6486}. Both their topological nature and the associated quantum phase transition elude the Landau paradigm. In addition to that, since the density of states vanishes along the nodal line, the Coulomb interaction between electrons is very weakly screened, leading to strong coupling. Indeed, strong electronic correlations and signatures of hydrodynamic behavior have been observed respectively in ZrSiSe \cite{Shao2020} and NbP \cite{Gooth2017}, and further confirmed by an extremely low value of the viscosity to entropy ratio \cite{PhysRevResearch.3.033003}. In a nutshell, the applicability of a weakly-coupled field theory description for certain TS is questionable (e.g., the breakdown of Fermi liquid theory in TS \cite{PhysRevB.98.205113}). In this regard, holographic methods, which have been already successfully applied to strongly-coupled electronic phases of matter \cite{zaanen2015holographic,hartnoll2018holographic,RevModPhys.95.011001}, present a viable alternative tool \cite{Landsteiner:2015lsa,Landsteiner:2015pdh,Landsteiner:2016stv} (see \cite{Landsteiner:2019kxb} for a review on holographic semimetals).

In Weyl semimetals, valence and conduction bands cross in single points, the Weyl nodes \cite{doi:10.1146/annurev-conmatphys-031016-025458,Jia2016}. In this case, the anomalous Hall effect, which appears as a consequence of the chiral anomaly \cite{Landsteiner:2016led}, has been early recognized as a non-local order parameter for the topological quantum phase transition (TQPT) between WSM and insulating states. This result, which is based on the proportionality between this transport coefficient and the distance between the Weyl nodes in wave-vector space, has been derived both using weakly-coupled field theory techniques \cite{Colladay:1998fq} and holographic methods at strong coupling \cite{Landsteiner:2015pdh}. In holographic WSM, quantum information quantities such as entanglement entropy \cite{Baggioli:2020cld} and butterfly velocity \cite{Baggioli:2018afg} have been shown to be successful probes for the TQPT. 

Nodal line semimetals \cite{fang2016topological} (e.g., Ca$_3$P$_2$~\cite{doi:10.1063/1.4926545,PhysRevB.93.205132}, PbTaSe$_2$ \cite{Bian2016}, ZrSiS \cite{Schoop2016}), in which the conduction and the valence bands touch along a one-dimensional curve in the three-dimensional Brillouin zone, represent an even more difficult challenge. Because of the absence of topologically protected surface states, the identification of a DC transport coefficient as an order parameter for the TQPT is not possible. A robust probe for nodal topology and the quantum phase transition in NLSM has not been found yet. For weakly-coupled NLSM, the power-law scaling of the shear viscosity in the collisionless limit has been proposed as a signature of nodal topology \cite{PhysRevB.101.161111}. More recently, using a holographic model for strongly coupled NLSM \cite{Liu:2018bye}, Ref.\cite{Rodgers:2021azg} showed that the DC electrical conductivity at low temperature displays a structure reminiscent of a quantum critical fan, which may provide a probe for the TQPT. Finally, in CaAgAs \cite{PhysRevResearch.2.012055},  quantum oscillations have been proven to be sensitive to the topology of the Fermi surface.

In this work, using holographic methods, we propose that the entanglement entropy, and quantities directly related to it, provide an order parameter for strongly-coupled NLSM, able to probe the nodal topology and the underlying topological quantum phase transition. 
Our work is motivated by the results for weakly-coupled NLSM presented in \cite{PhysRevB.95.235111}. 
In weakly-coupled NLSM, the gapless modes around the $1$-dimensional Fermi nodal line contribute dominantly to the entanglement entropy (EE) of the whole many-body system, while the contributions to EE from the gapped modes play only a subleading role. 
The EE from the gapless modes can further be computed via a generalized Widom formula. Following this procedure, the EE is found to be finite and proportional to the length of the Fermi nodal line \cite{PhysRevB.95.235111}. 
Therefore, the EE is able to probe the existence and the size of the $1$-dimensional Fermi nodal line, and it can be viewed as an order parameter for the topological phase transition between a topological NLSM phase and a trivial phase. 
Many of the features of EE are independent of the strength of the coupling, and in a sense they are universal. As an example, in short-range entangled systems, the leading term in the EE obeys the so-called area law \cite{Srednicki:1993im}, which is recovered in strongly coupled holography models \cite{Ryu:2006bv,Ryu:2006ef}. At the same time, area law violations in systems with $2$-dimensional Fermi surfaces appear in both  weakly-coupled \cite{PhysRevLett.96.100503,PhysRevLett.105.050502} and strongly-coupled systems \cite{Ogawa:2011bz,Huijse:2011ef}. 
Inspired by the promising results for weakly-coupled NLSM \cite{PhysRevB.95.235111}, we consider strongly-coupled NLSM described by the holographic model proposed in \cite{Liu:2020ymx}, and we investigate whether and how holographic EE is able to detect the topological quantum phase transition and the properties related to nodal topology.

This paper is organized as follows. 
In Sec.\ref{Sec:weak}, we review the weakly coupled field theory for NLSM and the behavior of entanglement entropy in NLSM using the generalized Widom formula proposed in \cite{PhysRevB.95.235111}. 
In Sec.\ref{Sec:holo}, we briefly introduce the holographic model for the strongly coupled NLSM, and demonstrate the topological quantum phase transition from the topological NLSM phase to a trivial phase across a quantum critical point. 
In Sec.\ref{Sec:result}, we study the holographic entanglement entropy across the topological phase transition and propose an order parameter from holographic entanglement entropy to characterize the topological phase transition. 
In Sec.\ref{Sec:conclusion}, we summarize our results and discuss the open questions. 
The computational details are outlined in appendices \ref{appA} and \ref{appB}. 

\section{Preliminaries}
\label{Sec:weak}

In this section we will first review the weakly coupled field theory for NLSM \cite{fang2016topological,PhysRevB.84.235126}, and then collect the known properties on the entanglement entropy in those systems.

\subsection{Weakly coupled nodal line semimetals}
\begin{figure}[th]
    \centering
    \includegraphics[width=0.8\linewidth]{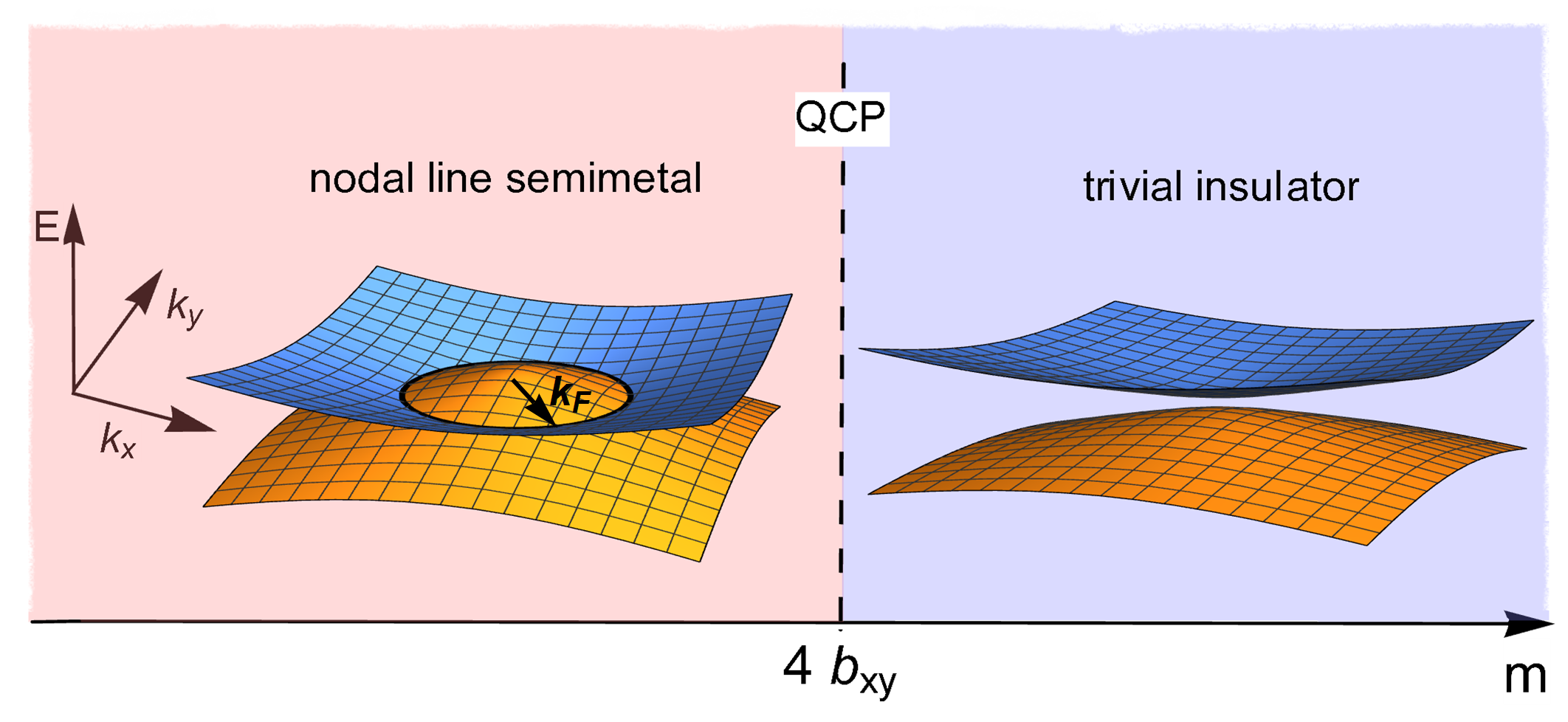}
    \caption{A cartoon of the topological quantum phase transition between a nodal line semimetal and a trivial insulating state. For simplicity, we set $k_z=0$ and we show only the lowest electronic bands. In the weakly-coupled theory, the quantum critical point is located at $4|b_{xy}|=|m|$.}
    \label{fig:1}
\end{figure}

A weakly-coupled field theory description for NLSM can be realized, in terms of a Dirac fermion $\psi$, using a $(3+1)$-dimensional Lorentz-violation Lagrangian \cite{fang2016topological,PhysRevB.84.235126},
\be
\mathcal{L}=\bar{\psi}\big(\gamma^\mu\partial_\mu-m-\gamma^{\mu\nu}b_{\mu\nu}+\gamma^{\mu\nu}\gamma^5b^5_{\mu\nu}\big)\psi\,.
\label{eqL:imNLSM}
\ee
Here,
\begin{equation}
    \bar{\psi}=\psi^\dagger i \gamma^0\,,\quad  \gamma^{\mu\nu}=\frac{i}{2}\left[\gamma^\mu,\gamma^\nu\right]\,,\quad \gamma^5=i\gamma^0\gamma^1\gamma^2\gamma^3\,,
\end{equation} 
and the anti-symmetric tensor operators obey a self-duality condition
\begin{equation}
    \bar{\psi}\gamma^{\mu \nu}\gamma^5\psi=-\frac{i}{2}\varepsilon^{\mu\nu}_{~~~\alpha\beta}\bar{\psi}\gamma^{\alpha \beta}\psi\,.
\end{equation}
As we turn on a background for the two-form field, $b_{xy}=-b_{yx}$, $b^5_{tz}=-b^5_{zt}=ib_{xy}$, a topological semimetal with a nodal line in the $k_z=0$ plane is realized. As long as $4|b_{xy}|>|m|$, the valence and conduction bands touch along a Weyl circle with radius 
\begin{equation}
k_F\equiv\sqrt{k_x^2+k_y^2}=\sqrt{16b^2_{xy}-m^2}\,.
\end{equation} 
At the critical value, $4|b_{xy}|=|m|$, the nodal circle collapses to a nodal point. The system at the quantum critical point is a Dirac semimetal. 
For $4|b_{xy}|<|m|$, the nodal topology disappears via a topological quantum phase transition, and the system is a trivial insulator. These three phases display different topological properties in their electronic spectrum on the $k_z=0$ plane. Fig.\ref{fig:1} provides a cartoon of the phase diagram for weakly coupled NLSM.\\

\subsection{Nodal entanglement entropy}

In this subsection, we review some known results for entanglement entropy (EE) in quantum field theories, in particular, for fermionic systems with nodal surfaces. Most of the analysis is taken from \cite{PhysRevB.95.235111}. 

Let us consider the ground state of a $(d+1)$-dimensional relativistic field theory, then, the entanglement Hamiltonian for a strip with width $l_{x_1}$ takes the form (Bisognano-Wichmann theorem \cite{doi:10.1063/1.522898,Bianchi:2012ev})
\bea
H=\int_{x_1=a}^{l_{x_1}}d^dx\left(2\pi x_1\right)\mathcal{H}\,,
\eea
where $a$ is the lattice size, representing the UV cutoff. 
The reduced density matrix is then given by
\bea
\rho\propto \text{exp}\left\{-\int_{x_1=a}^{l_{x_1}}d^dx\left(2\pi x_1
\right)\mathcal{H}\right\}\,,
\eea
where the local temperature $T(x)=1/(2\pi x_1)$ is defined. 
The dominant contribution to the entanglement entropy is given by the integral of the local thermal entropy densities:
\bea
S_{ent}=\int_{x_1=a}^{l_{x_1}}d^dx \,s_{th}\left(T(x)\right)\,.
\eea
For a $(d+1)$-dimensional many-body system with a $(d-q)$-dimensional Fermi surface, the thermal entropy density obeys the following relation:
\begin{equation}
    s(T)\propto \left(\frac{T}{v_F}\right)^q\,,
\end{equation}
where $v_F$ is the Fermi velocity, and the power is given by the codimension $q$. 
Therefore, the integral over the $x_1$ direction leads to:
\begin{align}
    & \text{log}\left(\frac{l_{x_1}}{a}\right)\quad \text{for}\quad q=1\,,\qquad \qquad \left(\frac{1}{a}-\frac{1}{l_{x_1}}\right)\quad \text{for}\quad q=2\,.
\end{align}

For a $(3+1)$-dimensional fermionic system with a Fermi surface of codimension $1$, e.g., a Fermi liquid, the entanglement entropy for a strip geometry takes the form
\be\label{ee1}
S\propto (k_F L)^{2} \log \frac{l}{a}+\dots\,,
\ee
where $k_F$ is the Fermi momentum, $L, l$ are the length and width of the strip respectively, $a$ is the lattice cutoff and the dots represent subleading terms \cite{PhysRevLett.96.100503,PhysRevLett.105.050502}. Equation \eqref{ee1} is also known as ``area law violation''. The realization of this violation in holographic systems can be found in \cite{Ogawa:2011bz,Huijse:2011ef}.

For a relativistic nodal line semimetal system, which is $(3+1)$-dimensional with a codimension $2$ Fermi nodal line having a unit tangent vector $\hat{t}$, the entanglement entropy can be written using the Widom formula for nodal lines \cite{PhysRevB.95.235111}:
\bea
\label{eq:EEweak}
S=N\left(\frac{1}{a}-\frac{1}{l}\right)\int_{P.S.}\int_{N.L.}|\hat{t}\times \hat{n}_r|^2\,,
\eea
with $N=9\zeta(3)/(64\pi^4)$. The inner integral is performed over the nodal line in the momentum space, while the outer integral is over the partitioning surface with normal vector $\hat{n}_r$. 

To continue, we consider a nodal circle with radius $k_F$, arising from a lattice structure with a lattice size $a$. 
Then, let us assume that the width of the strip is oriented along the $z$-direction, i.e. $l=l_z$, and the partitioning surface is on the $x$-$y$ plane. The nodal circle can reside on an arbitrary plane in momentum space, with $\theta$ representing the angle between the normal vectors of the nodal line in coordinate space and the $x$-$y$ plane. 
In this case, the entanglement entropy takes the following structure \cite{PhysRevB.95.235111}
\be\label{pp}
S\propto \left(1+\text{cos}^2\theta\right)k
_FL^2\left(\frac{1}{a}-\frac{1}{l_z}\right)+\dots\,.
\ee
Notice that the entanglement entropy reaches a maximum value at $\theta=0$ when the nodal line is located along the $k_x$-$k_y$ plane, and a minimum at $\theta=\pi/2$ when the nodal line is perpendicular to the $k_x$-$k_y$ plane. 
As evident from \eqref{pp}, the EE $S\propto k_F$ probes the size of the nodal circle, which results from the integral along the nodal line in momentum space. 
Remarkably, from the generalized Widom formula \eqref{eq:EEweak}, and after performing a summation of three entanglement entropies $S_i$ computed in three arbitrary entangled planes with three mutually orthogonal normal vectors $\{\hat{n}_1, \hat{n}_2, \hat{n}_3\}$, one obtains
\bea
\label{eq:EEunique}
S\equiv S_1+S_2+S_3=4\pi N k_F L^2\,\left(\frac{1}{a}-\frac{1}{l}\right)\,,
\eea
where we keep the strip length and width fixed when computing $S_i$, $l_i=l$. Eq.\eqref{eq:EEunique} provides a unique probe to distinguish NLSM from other anisotropic systems. 
Up to geometrical factors, the EE is linear in the radius of the nodal line $k_F$ and inversely proportional to the strip width $l$. This immediately implies that
\begin{equation}
    l^2\,\frac{\partial S(l)}{\partial l}\propto k_F\,\label{wwk}
\end{equation}
is independent of the UV cutoff, and an immediate probe for nodal topology.

\section{Holographic set-up}
\label{Sec:holo}
The holographic model for a $(3+1)$-dimensional strongly coupled NLSM \cite{Liu:2020ymx} is described by the following gravitational action 
\bea
\begin{split}
S&=\int d^5x\sqrt{-g}\bigg[R+12-\frac{1}{4}\mathcal{F}^2-\frac{1}{4}F^2
+\frac{\alpha}{3}\epsilon^{abcde}A_a \Big(3\mathcal{F}_{bc}\mathcal{F}_{de}+F_{bc}F_{de}\Big)
-(D_a \Phi)^*(D^a\Phi)\\&
-V_\Phi
-\frac{i}{6\eta}\epsilon^{abcde} \Big(B_{ab}H_{cde}^*-B_{ab}^* H_{cde}\Big)
-V_B
-\lambda|\Phi|^2B_{ab}^*B^{ab}\bigg]\,.
\end{split}
\label{eq:action}
\eea
The two gauge fields with strength $\mathcal{F}=dV$ and $F=dA$ correspond to vector and axial currents in the dual field theory. The bulk Chern-Simons term is chosen to ensure the correct anomalous Ward identity \cite{Landsteiner:2016led,Landsteiner:2019kxb}. The axially charged scalar field $\Phi$ plays the role of the mass term for the fermion $\psi$ in  \eqref{eqL:imNLSM}. The complex two form field $B_{ab}$ is dual to the rank-two operators described above. Its bulk field strength is defined as $H_{abc}=\partial_a B_{bc}+\partial_b B_{ca}+\partial_c B_{ab}-iq_2 A_a B_{bc}-iq_2 A_b B_{ca}-iq_2 A_c B_{ab}$. The Chern-Simons term for the two form field is introduced in order to impose the self-duality constraint.    
The potentials are chosen to be 
\bea
V_\Phi=m_1^2 |\Phi|^2+\frac{\lambda_1}{2} |\Phi|^4\,,\quad \text{and} \quad
V_B=m_2^2 B^*_{ab}B^{ab}+\frac{\lambda_2}{2}(B^*_{ab}B^{ab})^2\,.
\eea
Note that in the potential $V_B$ we have introduced an additional quartic term in comparison with the original model in \cite{Liu:2020ymx}. Upon tuning $\lambda_2$, we can control some of the IR properties related to the IR Lifshitz exponent in the NLSM phase \cite{Rodgers:2021azg}, which might render the holographic model more similar to real NLSM systems. For more details about the holographic model, see \cite{Liu:2020ymx}. The calculation details on the background of the system can be found in appendix \ref{appA}. 

The model can be specified by setting the following parameters $(\alpha, \eta, m_1^2, \lambda_1, q_1, q_2, m_2^2, \lambda_2,\\ \lambda)$. 
We choose $m_1^2=-3, m_2^2=1, \eta=2$ in order to match the conformal dimensions of the dual operators. 
Since we consider the zero density case, i.e. the gauge fields $V_a$ and $A_a$ vanish, the values of the parameters $q_1, q_2$, and $\alpha$ play no role in our analysis. 
The quartic term $\propto \lambda_2$ controls the Lifshitz exponent in the NLSM phase, which is independent of $\lambda_1$ and $\lambda$. For simplicity, we first choose $\lambda_2=0$ as an example and then discuss effects from general $\lambda_2$. 
For numerical convenience, we choose $\lambda_1=\lambda=2$ and $\lambda_2=0$.  We emphasize that the coupling $\propto \lambda$ in \eqref{eq:action} contributes to the effective mass of the scalar $\Phi$. In order to ensure the absence of any instability, we will always work with $\lambda>0$. \color{black}

In the following we will show the properties of the solutions at zero and finite temperature. 

\subsection{Zero temperature solutions}
\label{sssec:ir}
For the zero temperature solutions, we use the following ansatz
\bea
\label{eq:ztansatz}
\begin{split}
&ds^2=\frac{dr^2}{r^2}+u(-dt^2+dz^2)+f(dx^2+dy^2)\,,\\
&\Phi=\phi\,,\,~~B_{xy}=-B_{yx}=\mathcal{B}_{xy}\,,\,~~B_{tz}=-B_{zt}=i\mathcal{B}_{tz}\,,
\end{split}
\eea
where all the bulk fields $u, f, \phi, \mathcal{B}_{xy}, \mathcal{B}_{tz}$ are functions of the radial coordinate $r \in[0,\infty]$. 
The behaviors of the matter fields near the asymptotically AdS$_5$ boundary, $r\rightarrow \infty$, are given by
\be
\label{eq:dictionary}
\mathop{\text{lim}}_{r\rightarrow \infty}~r\phi=M\,,~~~~~~\mathop{\text{lim}}_{r\rightarrow \infty}~r^{-1}\mathcal{B}_{tz}=\mathop{\text{lim}}_{r\rightarrow \infty}~r^{-1}\mathcal{B}_{xy}=b\,,
\ee
where 
$M$ and $b$ represent external sources for the scalar operator $\bar{\psi}\psi$ and tensor operators $\bar{\psi}\gamma^{\mu\nu}\psi$ and $\bar{\psi}\gamma^{\mu\nu}\gamma^5\psi$ respectively. $M$ and $b$ play the role of the parameters $m$ and $b_{xy}$ in the weakly-coupled theory \eqref{eqL:imNLSM}.

At zero temperature, there exist three different phases with distinguished infrared (IR) ($r \rightarrow 0$) geometries, depending on the value of $M/b$.  
For $M/b<(M/b)_c\approx 1.1667$, the bulk scalar field $\phi$ vanishes in the IR. The near-horizon geometry enjoys an anisotropic scaling symmetry $
r^{-1}\rightarrow s^{\delta/\alpha}\,r^{-1}\,,(t,z)\rightarrow s^{\delta/\alpha}(t,z)\,, (x,y)\rightarrow s(x,y)\,$ 
with 
scaling exponent 
$
{\bf z}\equiv\delta/\alpha>1\,
$. The dual many-body system is in a strongly coupled NLSM phase. For $M/b=(M/b)_c$, the scalar field is finite while the two-form field obeys a power law in the IR, and the system is quantum critical. The near horizon geometry has the same form of anisotropic scaling symmetry but with a different exponent. When $M/b>(M/b)_c$, at leading order, in the IR the scalar field is finite while the two-form field vanishes. This IR solution obeys a relativistic scaling symmetry, i.e., ${\bf z}=1$. The dual field theory is in a topologically trivial phase. The location of the quantum critical point, $(M/b)_c$, depends on the couplings $\lambda, \lambda_1$ and $\lambda_2$ in the potentials. At finite temperature, the sharp quantum phase transition becomes a smooth crossover.  
\begin{figure}
    \centering
    \includegraphics[width=0.6\linewidth]{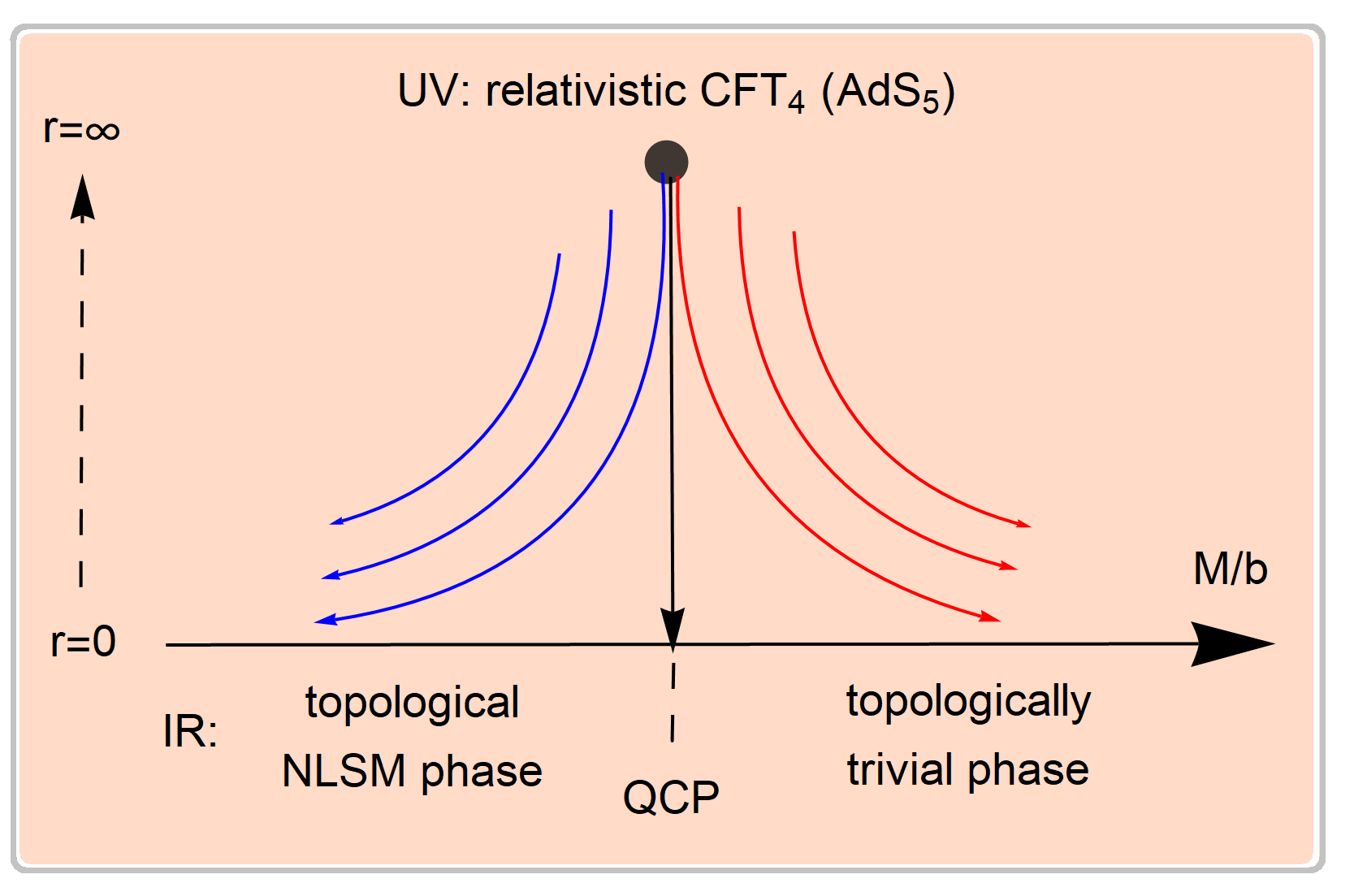}
    \caption{The $T=0$ RG flow and its geometric representation along the radial holographic extra-dimension.}
    \label{fig:rg}
\end{figure}
These three different phases, characterized by different IR geometries, exist depending on the value of $M/b$. A cartoon of the RG flow structure at $T=0$, and its geometrization, is shown in Fig.\ref{fig:rg}. Notice that the IR fixed points considered are stable under RG deformations, since they do not display any relevant direction. In other words, they are the real ground states of the system.

In the following, we introduce these three phases in detail.\\ 

\noindent
{\bf Topological NLSM phase:}~~ This phase exists below a critical value $M/b<(M/b)_c\approx 1.1667$.
In the IR limit $r\rightarrow 0$, at leading order, the solution reads
\bea
\label{eq:IRNLSM}
u=u_0r^\delta\,,~~~
f=f_0r^\alpha\,,~~~
\phi=0\,,~~~
\mathcal{B}_{tz}=u_0b_{tz0}r^\delta\,,~~~
\mathcal{B}_{xy}=f_0b_{xy0}r^\alpha\,. 
\eea
The parameters $\{\delta,\alpha, b_{tz0},b_{xy0}\}$ satisfy 
\bea
\begin{split}
&b_{tz0}=\sqrt{\frac{\alpha-\sqrt{\alpha^3\delta}}{2\lambda_2(\alpha-\delta)}}\,,\quad
b_{xy0}=\sqrt{\frac{\delta-\sqrt{\alpha\delta^3}}{2\lambda_2(\alpha-\delta)}}\,,\\
&\lambda_2=\frac{1-\alpha\delta}{-24+\alpha^2+4\alpha\delta+\delta^2}\,,\quad
0=(\alpha-\delta)^2+3\sqrt{\alpha\delta}\left((\alpha+\delta)^2-16\right)\,.\label{nnl}
\end{split}
\eea
From the above relations, we see that  without the quartic term in $V_B$, i.e. $\lambda_2=0$, the ratio $\delta/\alpha$ is completely fixed. Therefore the quartic term in $V_B$ is introduced to make the IR geometry more general, with a tunable scaling exponent. For given $\lambda_2$, all the four parameters are 
completely determined.  Interestingly, using the null energy condition (NEC), we have $\delta/\alpha\geq 1$. Moreover, we have $b_{tz0}/b_{xy0}=(\delta/\alpha)^{-1/2}$, which implies $b_{tz0}<b_{xy0}$. The IR geometry displays an emergent anisotropic scaling symmetry 
\be
\label{Eq:Lifshitzsymmetry}
r^{-1}\rightarrow s^{{\bf z}}\,r^{-1}\,,~~~(t,z)\rightarrow s^{{\bf z}}(t,z)\,,~~~ (x,y)\rightarrow s(x,y)\,\ee 
with scaling exponent 
\be
{\bf z}\equiv\frac{\delta}{\alpha}\,,
\ee
which measures the anisotropy of the ground state, and is always larger than $1$. This type of IR solution corresponds to the topological NLSM phase.  \\

\noindent
{\bf Quantum critical point (QCP):}~~ The QCP is located at $M/b=(M/b)_c$. The solution, at leading order, is given by
\bea
\begin{split}
\label{eq:IRQCP}
u&=u_c\,r^{\delta}\,,\quad
f=f_c\,r^{\alpha}\,,\quad
\phi=\phi_c=\sqrt{\frac{3+2\lambda(b_{tzc}^2-b_{xyc}^2)}{\lambda_1}}\,,\\
\mathcal{B}_{tz}&=u_c\, b_{tzc}r^{\delta}\,,\quad
\mathcal{B}_{xy}=f_c\, b_{xyc}r^{\alpha}\,.
\end{split}
\eea
The parameters $\delta,\alpha$ are functions of $\lambda,\lambda_1, \lambda_2$ which can be found by solving the following algebraic equations:
\bea
\begin{split}
0&=\alpha\,b_{xyc}+2\lambda_2b_{tzc}^3-(1+2\lambda_2 b_{xyc}^2+\lambda\phi_c^2)b_{tzc}\,,\\
0&=\delta\,b_{tzc}-2\lambda_2 b_{xyc}^3 -(1-2\lambda_2 b_{tzc}^2+\lambda\phi_c^2)b_{xyc}\,,\\
0&=\alpha^2 - \delta^2 - 8 \lambda_2 b_{tzc}^4  + 8 \lambda_2 b_{xyc}^4 + 
 4(1 + \lambda \phi_c^2) (b_{tzc}^2+b_{xyc}^2) \,,\\
0&=\alpha^2 + \alpha \delta + \delta^2 -12 - 2 \lambda_2 (b_{tzc}^2 -  b_{xyc}^2)^2 \
- 3 \phi_c^2 + \frac{\lambda_1 \phi_c^4}{2}\,.
\end{split}
\eea
This type of IR solution corresponds to the quantum critical point.\\

\begin{figure}[h!]
  \centering
\includegraphics[width=0.45\textwidth]{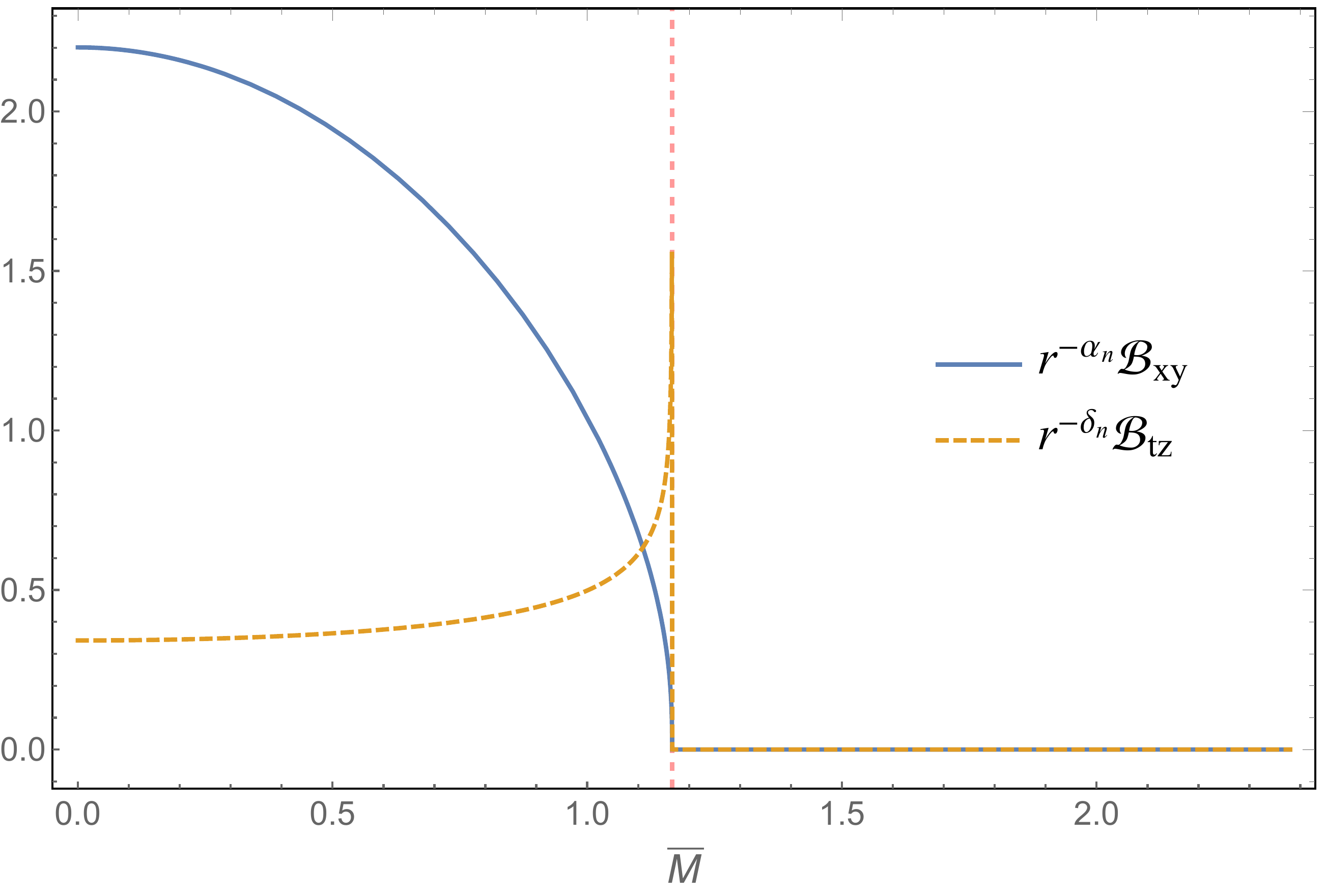}~~\quad 
\includegraphics[width=0.45\textwidth]{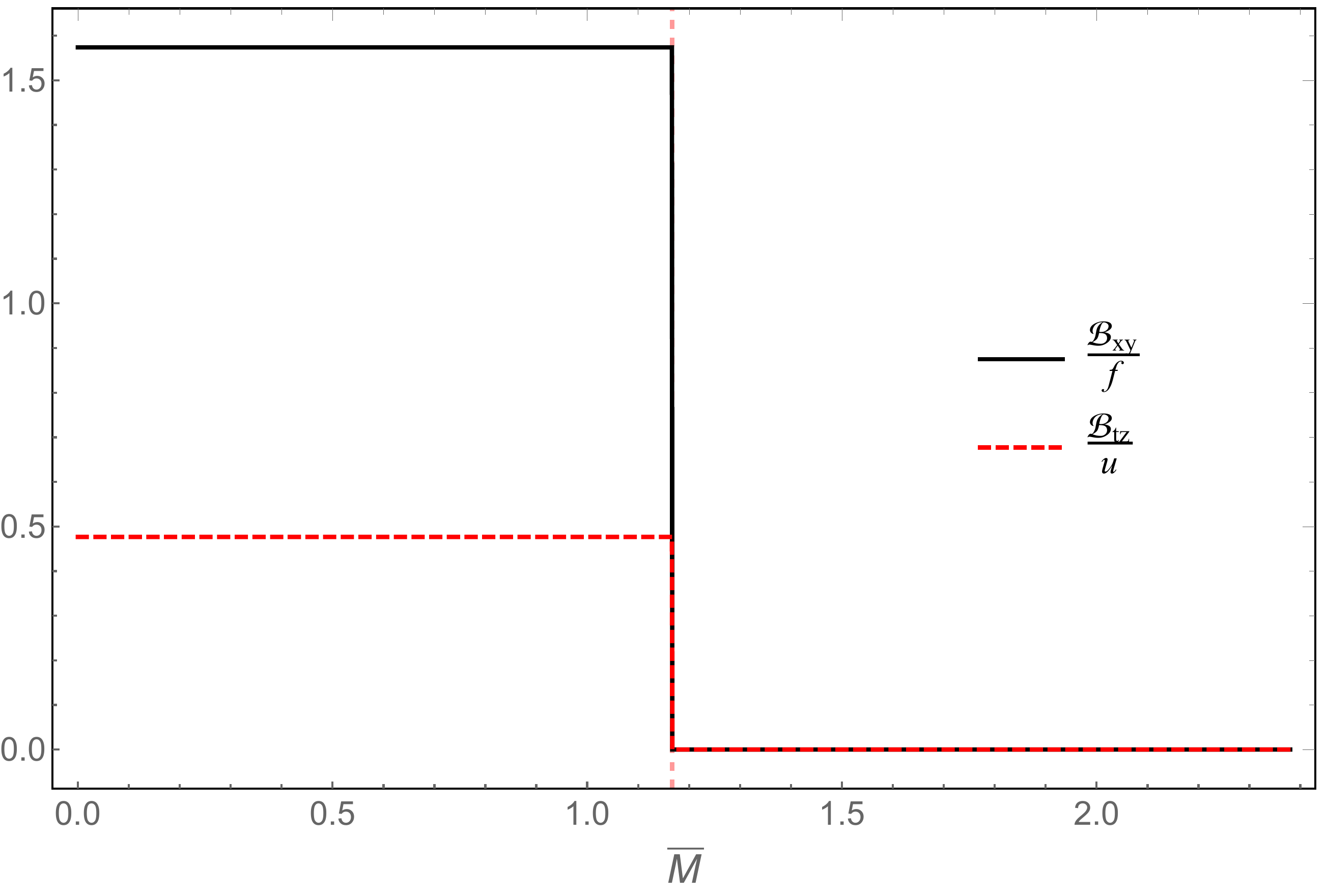}
  \caption{\small \textbf{Left: } $r^{-\alpha_n} \mathcal{B}_{xy}  $, $r^{-\delta_n} \mathcal{B}_{tz}  $ and $f(r)$ in the deep IR  limit as a function of $M/b$. \textbf{Right:} The ratios between the components of the two-form bulk field and the metric components in the near-horizon limit. The vertical dashed line locates the quantum critical point separating the NLSM phase ($M/b \lessapprox 1.1667$) and the trivial phase ($M/b \gtrapprox 1.1667$).}
  \label{fig:profile}
\end{figure}

\noindent
{\bf Topologically trivial phase:}~~ This phase is found for $M/b>(M/b)_c$. At leading order, the near-horizon solution is given by
\bea
\label{eq:IRTrivial}
\begin{split}
u=u_0r^{\delta}\,,\quad
f=f_0r^{\alpha}\,,\quad
\phi=\phi_0\,,\quad
\mathcal{B}_{tz}=0\,,\quad
\mathcal{B}_{xy}=0\,.
\end{split}
\eea
From the equations of motion, we have 
\bea
\phi_0=\sqrt{\frac{3}{\lambda_1}}\,,\quad
\delta=\alpha=\sqrt{\frac{3+8\lambda_1}{2\lambda_1}}\,.
\eea
In the topologically trivial phase, the low energy system retains the relativistic scaling symmetry with  ${\bf z}\equiv\delta/\alpha=1$. In other words, the near-horizon geometry in the deep IR is AdS$_5$. \\
In summary, the aforementioned exponents related to the IR solutions, and corresponding to the different phases in the dual field theory, are outlined in table \ref{table:parameters}. For later convenience, we define the scaling parameters $\delta$ and $\alpha$ in the NLSM phase, obtained from \eqref{nnl}, as $\delta_n,\alpha_n$.
The renormalized quantities $r^{-\alpha_n} \mathcal{B}_{xy}  $ and $r^{-\delta_n} \mathcal{B}_{tz}$ in the deep IR ($r\rightarrow 0$) 
depend on $M/b$ and are shown in Fig.\ref{fig:profile} across the whole phase diagram. 
\begin{table}[h]
\centering
\setlength{\tabcolsep}{7mm}{
\begin{tabular}{|c|c|c|c|c|}
\hline
 & $\delta$ & $\alpha$ & ${\bf z}\equiv \delta/\alpha$ & $1/{\bf z}$ \\ \hline
NLSM phase& 3.303 & 0.303 & 10.908 & 0.092 \\ \hline
Critical phase& 2.99 & 1.008 & 2.968 & 0.337 \\ \hline
Trivial phase& 2.179 & 2.179 & 1 & 1 \\ \hline
\end{tabular}}
\caption{Scaling exponents for the near-horizon IR geometries.}
\label{table:parameters}
\end{table}
\\

Both quantities display a critical behavior across the quantum critical point. In particular, they are zero in the trivial phase, $M/b>(M/b)_c$, and finite in the topological NLSM phase, reaching a constant value at $M/b=0$. $r^{-\alpha_n} \mathcal{B}_{xy}$ shows a continuous behavior across the critical point which can be extracted numerically and it is given by:
\begin{equation}
   r^{-\alpha_n} \mathcal{B}_{xy} \propto (\bar{M}_c-\bar{M})^{0.37}\,.
\end{equation}
Interestingly, the critical behavior of these two quantities is exactly the same of the metric components $r^{-\alpha_n}f(r),r^{-\delta_n}u(r)$, as shown in the right panel of Fig.\ref{fig:profile}. The ratios $B_{xy}/f$ and $B_{tz}/u$, computed at $r \rightarrow 0$, are indeed zero in the trivial phase and constant in the NLSM phase. These combinations correspond to nothing else than $B_\mu^{~\nu}$, which jumps discontinuously across the QCP. The value of $B_x^{~y}$ and $B_t^{~z}$ in the NLSM phase is a function of the anisotropic exponent ${\bf z}$. The exponent ${\bf z}$ controls the breaking of the SO(3) symmetry down to SO(2). In the trivial phase ${\bf z}=1$, and the IR emergent field theory has a relativistic scaling represented by an AdS$_5$ type IR geometry. In the NLSM phase, ${\bf z}$ depends on the parameter $\lambda_2$,  and at the quantum critical point, ${\bf z}$ depends on the parameters $\lambda, \lambda_1$ and $\lambda_2$. Importantly, the null energy condition implies ${\bf z}\geq 1$ for both ${\bf z}_{\text{NLSM}}$ and ${\bf z}_c$ (where the label $c$ stands for ``critical'') \cite{PhysRevD.82.084002}. In general,  ${\bf z}_{\text{NLSM}}>{\bf z}_c>{\bf z}_{\text{trivial}}=1$ is a universal relation independent of the parameters of the model. 

Based on the properties of the dual fermionic spectral function, it has been shown that in the NLSM phase there exist multiple nodal line Fermi surfaces which are topologically stable \cite{Liu:2020ymx}. This confirms that the holographic model describes a topological quantum phase transition between a NLSM and a trivial phase (see also \cite{Rodgers:2021azg} for a recent analysis of the thermodynamics and transport properties of a similar holographic NLSM model).

\subsubsection{The background geometry at zero temperature}

In the previous section, we have studied the deep IR limit ($r \rightarrow 0$) of the zero-temperature solutions. Here, we show the behavior of the whole geometry for different values of $M/b$. 

\begin{figure}[h!]
  \centering
\includegraphics[width=0.32\textwidth]{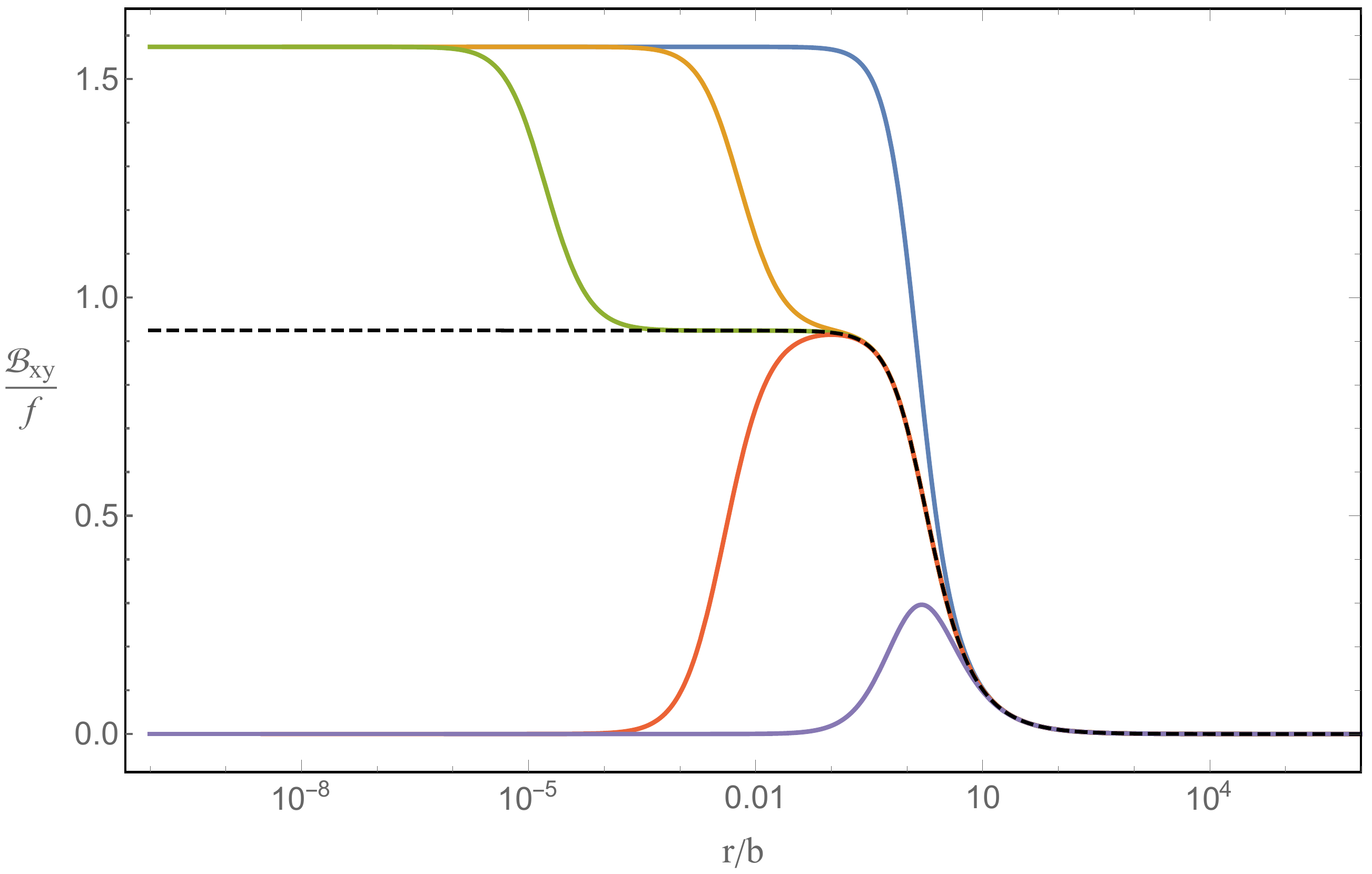}
\includegraphics[width=0.32\textwidth]{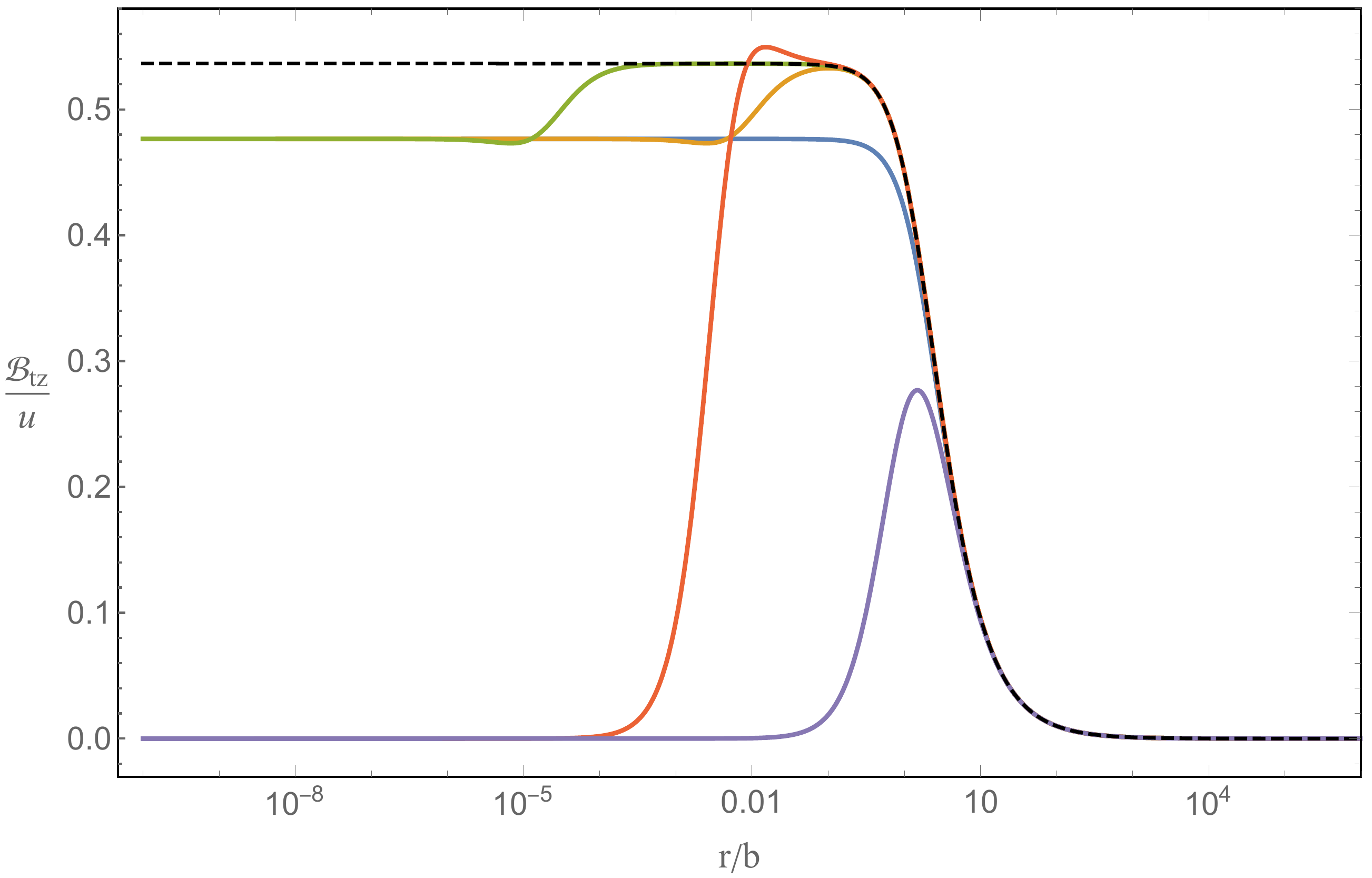}
\includegraphics[width=0.312\textwidth]{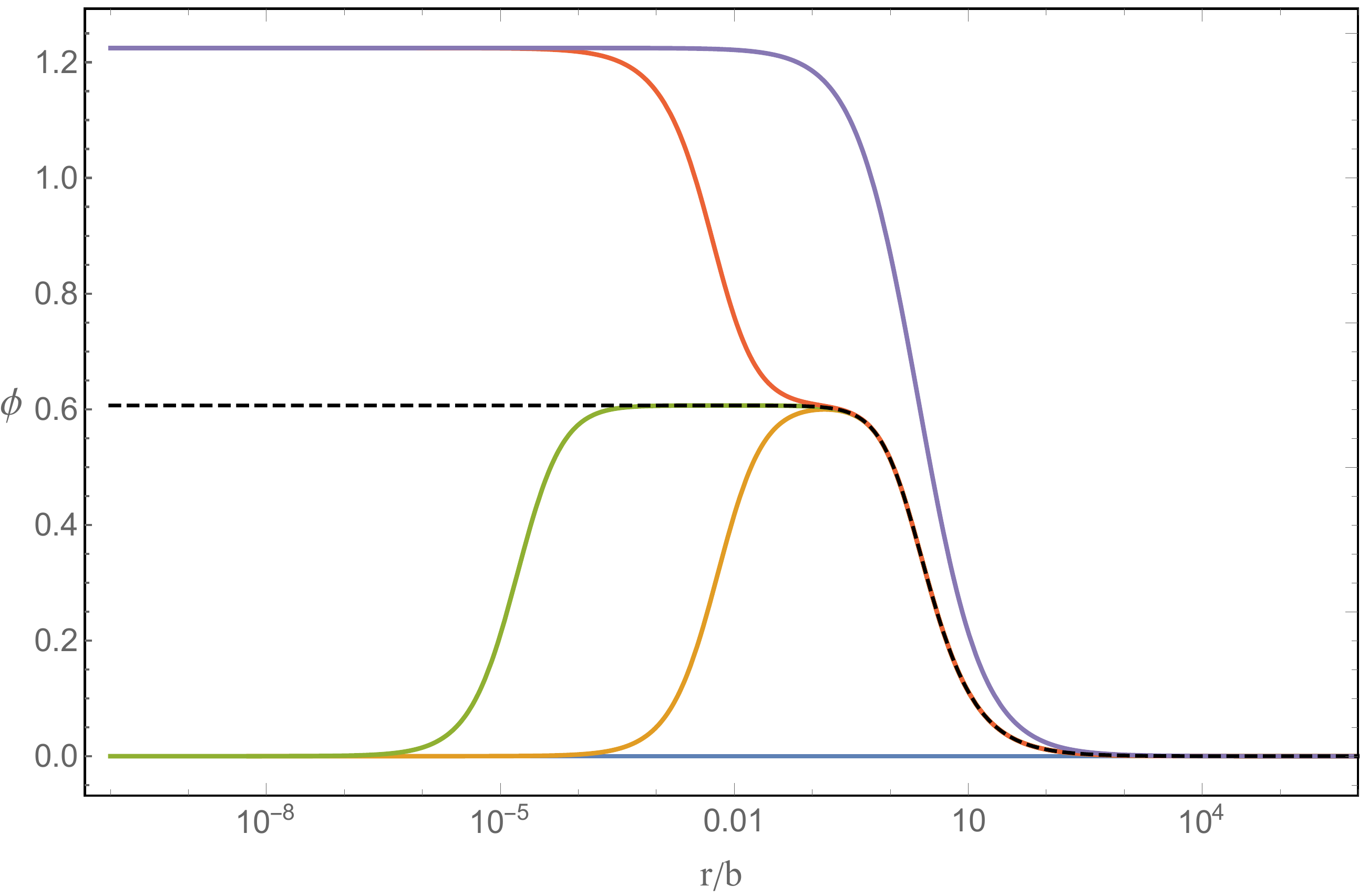}
  \caption{\small 
  The radial profiles for the different metric/matter fields for different values of the external parameter $M/b$ at zero temperature: $\mathcal{B}_{xy}/f$ ({\bf left}), $\mathcal{B}_{tz}/u$ ({\bf center}) and $\phi$ ({\bf right}). The different colors correspond to
  $\bar{M}=0$ (blue), $1.1666$ (orange), $\bar{M}_c-\delta$ (green), $\bar{M}_c$ (black dashed), $1.1668$ (red), $2.36$ (violet). The black dashed line is the RG flow ending in the IR quantum critical point.}
  \label{fig:profiles}
\end{figure}

Fig.\ref{fig:profiles} shows the profiles of the metric and matter fields along the radial direction across the quantum phase transition, including the configurations  
in the vicinity of the QCP. When $\bar{M}=0$, the scalar field $\phi$ vanishes. As $\bar{M}$ increases, and it is kept smaller than the critical value $\bar{M}_c$, the IR geometry is approached for smaller values of the renormalized radial coordinate $r/b$. 
More interestingly, in the vicinity of the QCP, i.e. $\bar{M} \approx \bar{M}_c$ (both below and above it), the geometry first flows to an intermediate scaling regime determined by the  critical state, and only afterward reaches the deep IR geometry.  

Physically, the radial profile of the different fields geometrizes the renormalization group (RG) flow of the dual field theory from the UV relativistic fixed point at $r=\infty$ to the IR fixed point at $r=0$ (see Fig.\ref{fig:rg} for a cartoon). Whenever the value of $M/b$ is close to the critical one, the RG flow trajectory passes very close to the quantum critical fixed point, and therefore all the physical properties reflect the nature of the QCP for an intermediate energy (or equivalently radial coordinate) range. The closer $M/b$ to the critical value, the larger the energy range in which the features of the QCP are observed. At exactly the critical value, the QCP is the IR endpoint of the RG flow.

\subsection{Finite temperature solutions}

At finite temperature, $T>0$, the holographic dual of the thermal state is an anisotropic black hole described by 
the following ansatz
\bea
\label{eq:bgfiniteT}
ds^2=-u\,dt^2+\frac{1}{u}dr^2+f(dx^2+dy^2)+h\,dz^2\,,
\eea
where the horizon locates at $r=r_0$, with $u(r_0)=0$. 

To obtain the bulk solutions, we first derive the near-horizon expansion for the bulk fields which are given by
\bea
\begin{split}
u&=4\pi T(r-r_0)+u_2(r-r_0)^2+\dots\,,\\
f&=f_0+f_1(r-r_0)+\dots\,,\\
h&=h_0+h_1(r-r_0)+\dots\,,\\
\mathcal{B}_{tz}&=b_{tz1}(r-r_0)+\dots\,,\\
\mathcal{B}_{xy}&=b_{xy0}+b_{xy1}(r-r_0)+\dots\,,\\
\phi&=\phi_0+\phi_1(r-r_0)+\dots\,.
\end{split}
\eea
The sub-leading parameters $\{u_2,f_1,h_1,b_{tz1},b_{xy1},\phi_1\}$ are functions of the leading order parameters $\{u_1=4\pi T,f_0,h_0,b_{xy0},\phi_0\}$ in the near horizon expansions. 
Using the scaling symmetries described in the previous sections, there remain only two free shooting parameters corresponding to the two dimensionless variables in the boundary field theory, i.e., $T/b, M/b$. 
By employing a shooting technique, we can construct numerically the profiles of the bulk fields at finite temperature.

\begin{figure}[h!]
  \centering
\includegraphics[width=0.5\textwidth]{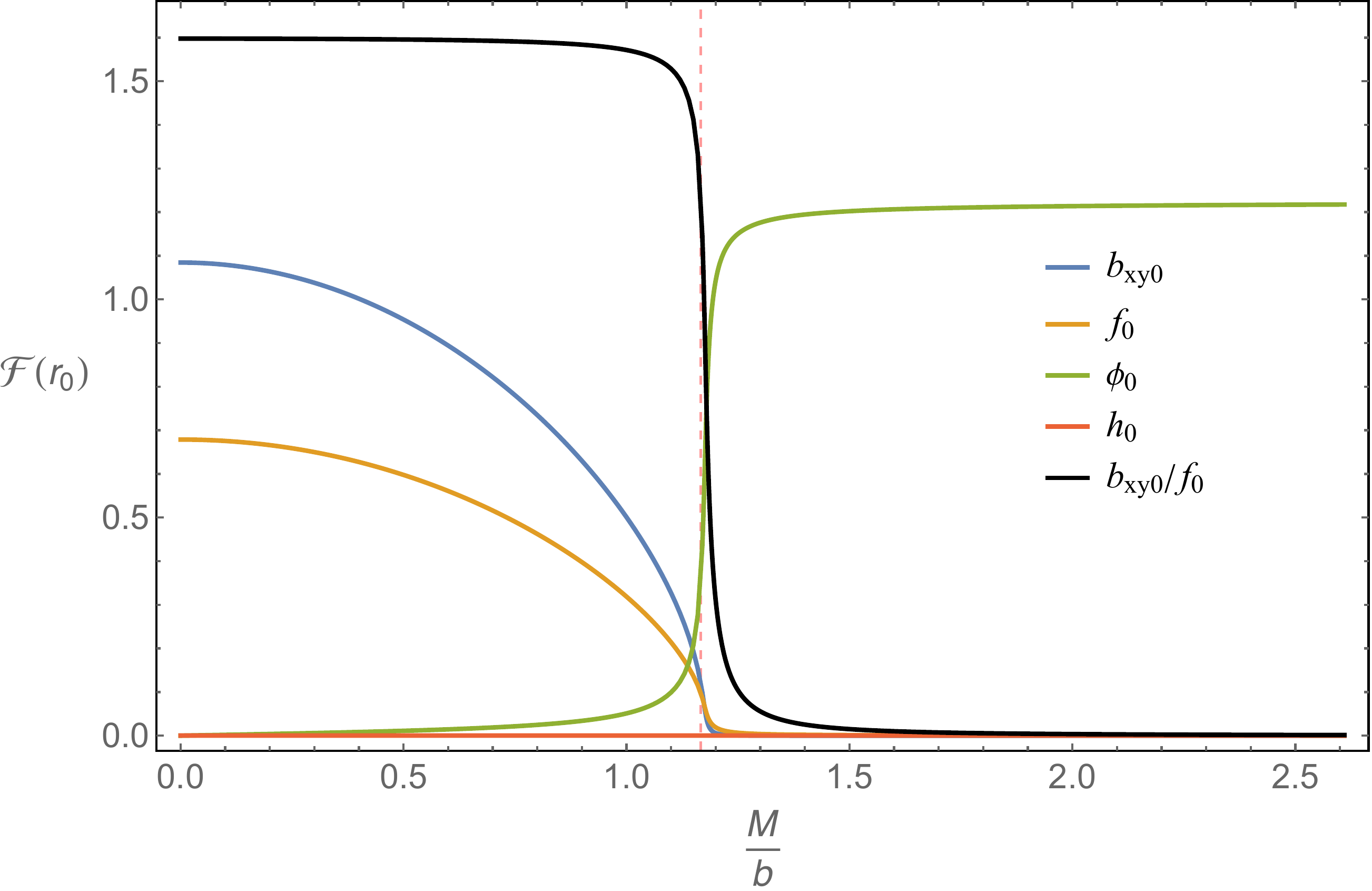}
  \caption{\small The bulk fields $\mathcal{F}=\{\mathcal{B}_{xy}, f, \phi, h, \mathcal{B}_{xy}/f\}$ computed at the horizon $r_0$ for $T/b=0.005$. The behavior of the horizon data as a function of $M/b$ is consistent with the expectation that 
  the quantum phase transition becomes a crossover at finite $T$.}
  \label{fig:horizon}
\end{figure}

At finite but low temperatures, the near-horizon geometries partly retain the zero-temperature properties. The topological quantum phase transition becomes a smooth crossover, as illustrated in Fig.\ref{fig:horizon} using the near horizon value of the fields. 
The horizon data $b_{xy0}$ and $f_0$ shrink to $0$ smoothly across the crossover, while their ratio remains constant, $b_{xy0}/f_0\simeq 1.60$, in the NLSM phase and decreases sharply to $0$ in the quantum critical region. 
At the same time, $\phi_0$ increases from $0$ to a finite constant across the crossover. For $\bar{M}\gg 1$, $f_0=h_0$ and isotropy is recovered. On the contrary, as $\bar{M}\rightarrow 0$, we have $f_0 \gg h_0$, implying a very strong anisotropy. At finite temperature, the interpolation between these two extremes is continuous. At low temperature, the black hole interiors show some ``topological'' features in the topologically NLSM phase \cite{Gao:2023zbd}. 

\section{Methods and results}
\label{Sec:result}

In this section, we first study the entanglement entropy of the holographic model introduced in the previous section, and then construct physical quantities to characterize the topological phase transition. 

\subsection{Holographic entanglement entropy
}\label{s3}
In order to compute the entanglement entropy of the strongly coupled nodal line semimetal systems from holography, we use the Ryu-Takayanagi formula \cite{Ryu:2006bv}. We consider the following 5-dimensional metric
\bea
ds^2=-g_{tt}\, dt^2+g_{rr}\,dr^2+g_{xx}dx^2+g_{yy}\,dy^2+g_{zz}\,dz^2\,,
\eea
where all the metric components are functions of the radial direction $r$. This general ansatz includes both the $T=0$ case, Eq.\eqref{eq:ztansatz}, and the finite $T$ one, Eq.\eqref{eq:bgfiniteT}. 

We focus on strip geometries. 
The first case is that of a strip aligned along the $x$ direction and defined by the interval
\begin{equation}
   x\in (-l_x/2, l_x/2)\,,\quad  y,z\in (-L/2, L/2)\,,\qquad L\rightarrow \infty\,.
\end{equation}
The extremal surface $\gamma$ is parameterized as $r=r(x)$. The induced metric on $\gamma$ is 
\be
ds_\gamma^2=\big( g_{rr}\,r'^2+g_{xx}\big)\,dx^2+g_{yy}\,dy^2+g_{zz} \,dz^2\,,
\ee
where $r'\equiv \partial_x r$. 
Following these notations, the area functional is given by
\be
\label{eq:arefun}
A=L^2 \int_{-l_x/2}^{l_x/2} dx \sqrt{g_{yy} g_{zz}\,\big( g_{rr}\,r'^2+g_{xx}\big)} \,.
\ee

From the variation of the area functional, we could obtain the equations for $r(x)$ whose solution gives the extremal surface. A more convenient way to reach the same result is as follows. Since the area functional does not depend on the coordinate $x$, there is an associated conserved quantity given by
\be\label{eq:coneq}
\frac{g_{xx}\sqrt{g_{yy}g_{zz}}}{\sqrt{g_{rr}r'^2+g_{xx}}}=\mathfrak{C}_1\,,
\ee
where $\mathfrak{C}_1$ is a constant. From \eqref{eq:coneq}, we have 
\be\label{eq:c1def}
\mathfrak{C}_1=\sqrt{g_{xx}g_{yy}g_{zz}}\Big{|}_{r=r_t}\,,
\ee
where $r_t$ indicates the location of the turning point given by the condition $r'=0$. Using the above identities, we get
\be\label{eq:eqrt}
\int_{r_t}^\infty dr \frac{\mathfrak{C}_1}{\sqrt{g^{rr}g_{xx}(g_{xx}g_{yy}g_{zz}-\mathfrak{C}_1^2)}}=\frac{l_x}{2}\,.
\ee
This equation has to be understood as an implicit equation giving the location of the turning point $r_t$. Then, using equation \eqref{eq:coneq}, the profile of the extremal surface $r(x)$ can be obtained. Notice that there might be cases for which multiple turning points exit. If that happens, one should identify the turning point corresponding to the minimal area for the extremal surface. 

Using the Ryu-Takayanagi prescription \cite{Ryu:2006bv}, the entanglement entropy is finally given by 
\be
S_x=\frac{L^2}{4G}\,\int_{-l_x/2}^{l_x/2} dx\,\frac{g_{xx}g_{yy}g_{zz}}{\mathfrak{C}_1}\,.
\ee
Notice that because of the SO(2) symmetry in the $x$-$y$ plane, $S_x=S_y$, where $S_y$ indicates the EE for a strip aligned along the $y$ direction.

\begin{figure}[h!]
  \centering
\includegraphics[width=0.45\textwidth]{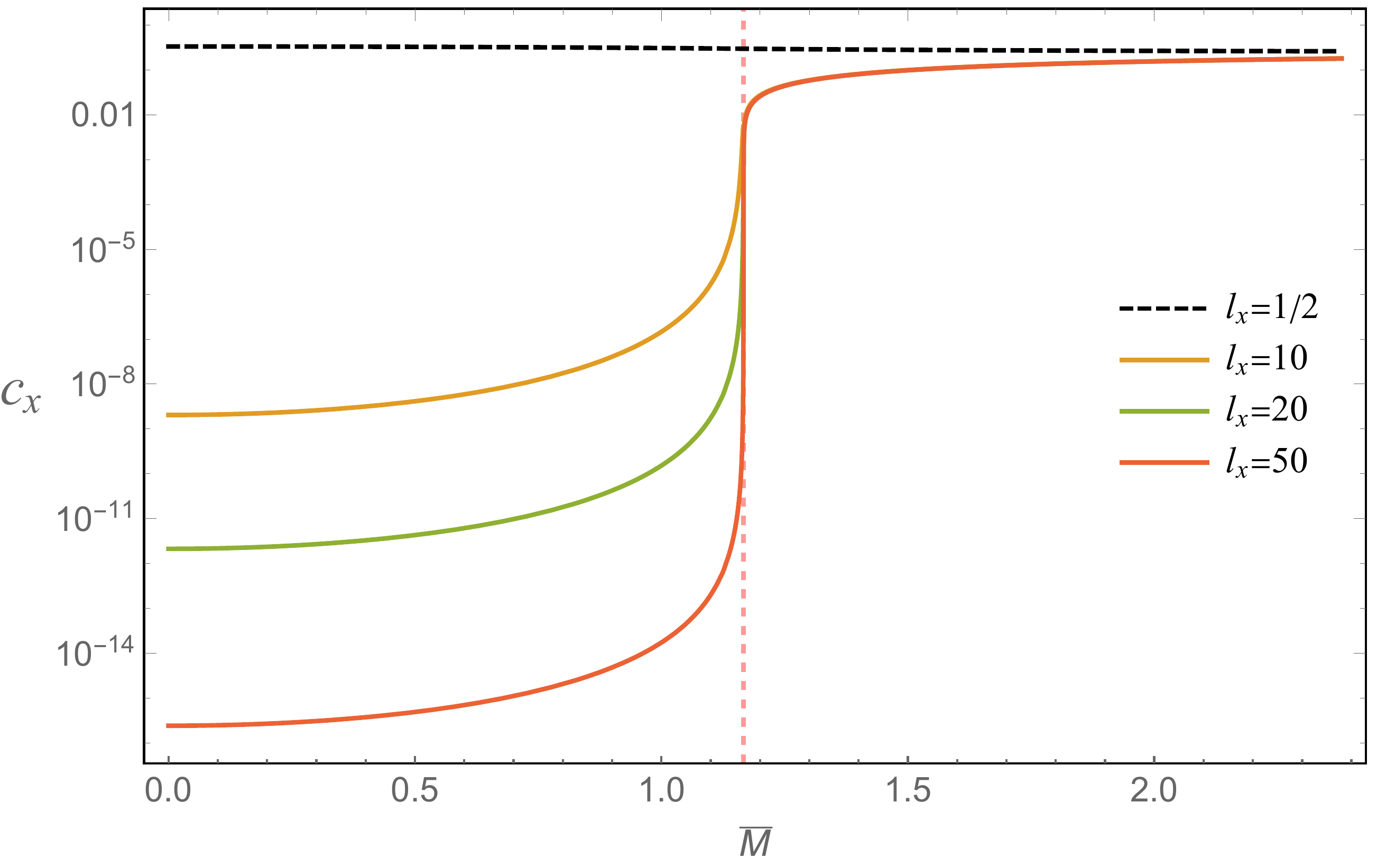}
\hspace{0.1pt}
\includegraphics[width=0.45\textwidth]{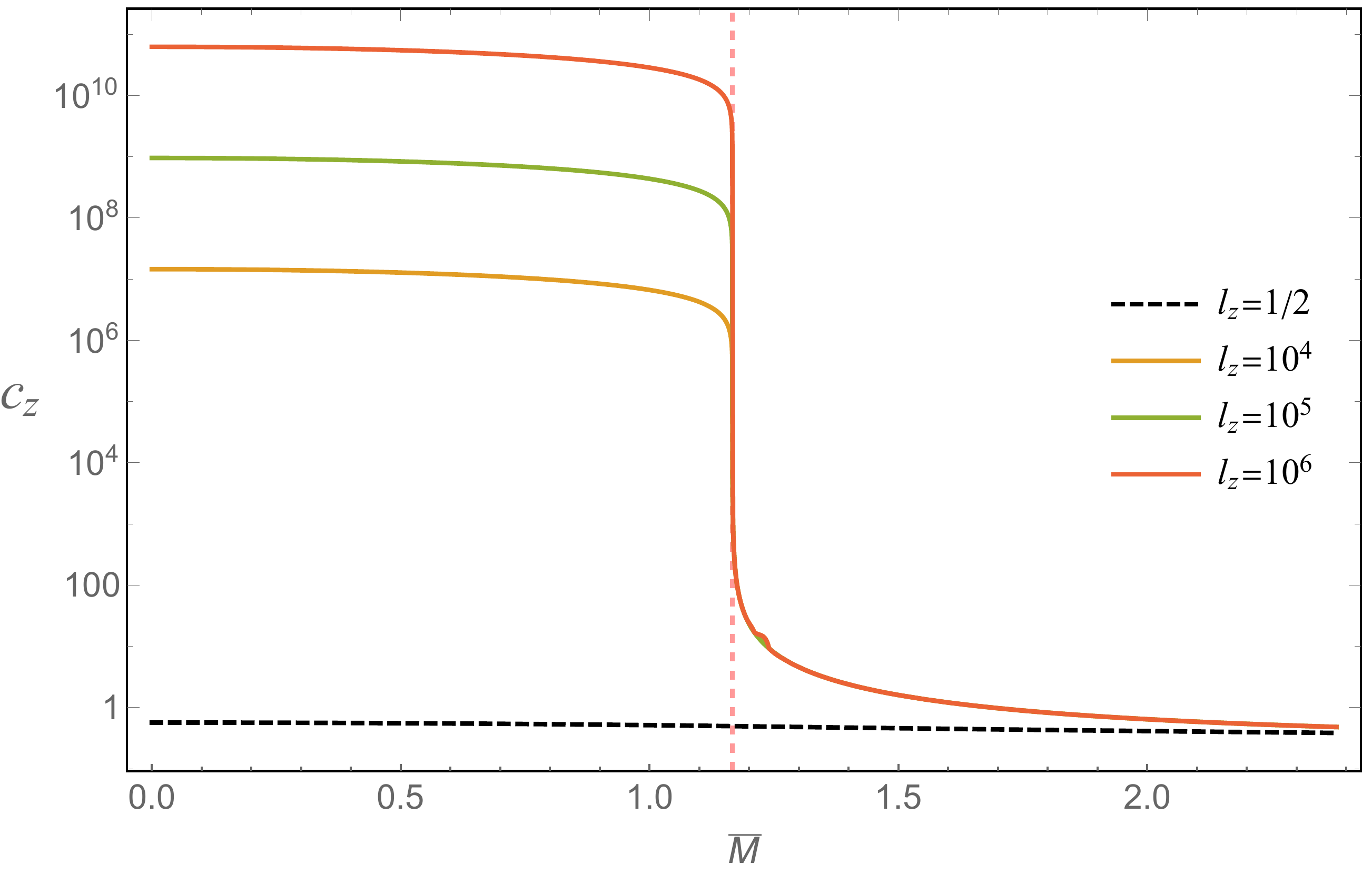}
  \caption{\small Logarithmic plots of $c_x$ ({\bf left}) and $c_z$ ({\bf right}) as a function of $\bar{M}=M/b$ in the large $l$ limit and for $T/b=0$.
  }
  \label{fig:clogplot}
\end{figure}

Taking advantage of Eq.\eqref{eq:coneq}, in the UV limit $r\to \infty$, we have
\be
x=\frac{l_x}{2}-\frac{\mathfrak{C}_1}{4r^4}+\dots\,.
\ee
Following the discussions in \cite{Myers:2012ed,Liu:2013una}, we assume that  the cutoff $r(\frac{l_x-\epsilon}{2},l)=r_c$ is fixed and we perform the variation of \eqref{eq:arefun} with respect to the strip length $l_x$. Then, we obtain
\be
\frac{4G}{L^2}\frac{\partial S_x}{\partial l_x}
=-\mathfrak{C}_1\,\frac{1}{r'}\frac{\partial r}{\partial l_x}\bigg{|}_{x=\frac{l_x-\epsilon}{2}}=\frac{1}{2}\mathfrak{C}_1\,.
\ee
 As a second case, we consider the strip along the anisotropic $z$ direction:
\begin{equation}
    z\in (-l_z/2, l_z/2)\,,\quad  x,y\in (-L/2, L/2)\,,\qquad L\rightarrow \infty\,.
\end{equation}
Now, the entanglement surface is parameterized by $r(z)$ and satisfies
\be
\label{eq:coneq2}
\frac{g_{zz}\sqrt{g_{xx}g_{yy}}}{\sqrt{g_{rr}{r^*}^2+g_{zz}}}=\mathfrak{C}_3\,,
\ee
where, $r^*\equiv \partial_z r(z)$, and
\be 
\mathfrak{C}_3=\sqrt{g_{xx}g_{yy}g_{zz}}\Big{|}_{r=r_t}\,,\ee 
with $r_t$ determined by 
\be
\label{eq:eqrt2}
\int_{r_t}^\infty dr \frac{\mathfrak{C}_3}{\sqrt{g^{rr}g_{zz}(g_{xx}g_{yy}g_{zz}-\mathfrak{C}_3^2)}}=\frac{l_z}{2}\,.
\ee
In this case, the entanglement entropy is 
\be
S_z=\frac{L^2}{4G} \, \int_{-l_z/2}^{l_z/2} dx\,\frac{g_{xx}g_{yy}g_{zz}}{\mathfrak{C}_3}\,.
\ee
Using similar arguments, we obtain 
\be
\frac{4G}{L^2}\frac{\partial S_z}{\partial l_z}
=\frac{1}{2}\mathfrak{C}_3\,.
\ee
Summarizing all the results above, we have 
\be
\label{eq:c-functions}
c_x=\frac{1}{2}\mathfrak{C}_1l_x^3\,,
~~~ 
c_y=\frac{1}{2}\mathfrak{C}_2l_y^3\,,
~~~
c_z=\frac{1}{2}\mathfrak{C}_3l_z^3\,,
\ee 
where
\begin{equation}
    c_i\equiv 4 G \frac{l_i^3}{L^2}\frac{\partial S_i}{\partial l_i}\,,\label{eq:defci}
\end{equation} 
and $\mathfrak{C}_i=\sqrt{g_{xx}g_{yy}g_{zz}}\Big{|}_{r=r_t}$ with $r_t$ determined by \eqref{eq:eqrt} for $i=x\,,y$, or \eqref{eq:eqrt2} for $i=z$. The behavior of these ``$c$-functions'' at zero temperature can be found in Fig.\ref{fig:clogplot}. We observe that both $c$-functions display a sharp feature at the quantum critical point $\bar{M}=\bar{M}_c$, which becomes more and more pronounced in the $l_i \rightarrow \infty$ limit. Intuitively, this can be understood as follows. In the limit $l_i \rightarrow \infty$, the extremal surface probe more and more the IR region of the bulk geometry and becomes more sensitive to the IR physics, which is crucial in the determination of the quantum critical point. Following this argument, the $c$-functions will be used to construct an order parameter able to probe the nodal topology and the quantum critical phase transition. Let us notice that in the NLSM phase, $c_x=c_y \ll c_z$. As a consequence, it would suffice to consider the $c$-function in the $z$ direction to construct a reliable order parameter.

\subsubsection{Entanglement entropy across the phase diagram}
In the following, we further discuss the properties of the entanglement entropy. The holographic entanglement entropies $S_x$ and $S_z$ display the following structure 
\bea
\label{eq:eeads5}
\begin{split}
\frac{4G}{L^2}\,S_x=\frac{r_c^2}{2}-\frac{M^2}{6}\text{ln}(r_c)+s_x\,,\quad
\frac{4G}{L^2}\, S_z=\frac{r_c^2}{2}-\left(\frac{M^2}{6}-b^2\right)\text{ln}(r_c)+s_z\,,
\end{split}
\eea
where $r_c^{-1}\ll l\ll L$, and 
$r_c$ is the cutoff near the AdS boundary  which can be identified as the UV cutoff in the dual field theory. In \eqref{eq:eeads5}, the first two terms come from the UV divergent structure, with the first one being a result of asymptotic AdS geometry \cite{Ryu:2006ef}, and the second one comes from the deformations on the UV CFT$_4$. These terms can be obtained analytically from the UV asymptotic expansion of fields in  \eqref{eq:uv}. The different coefficient in front of the subleading log-term in \eqref{eq:eeads5} is due to the anisotropic deformation in the UV CFT$_4$.  On the contrary, the terms 
 $s_x,s_z$ are finite in the limit $r_c\to \infty$, and they are not a feature of the UV fixed point. To simplify the notations, from now on, we will use the renormalized quantities $$
\bar{M}=\frac{M}{b}\,,\quad
\bar{T}=\frac{T}{b}\,,\quad
\bar{l}=b\,l\,,$$
such that the finite part of the entanglement entropies will be functions of the type 
\be 
s_{x,z}=s_{x,z}(r_c, \bar{M},\bar{T},\bar{l}_{x,z})\,.
\ee 
We shall discuss $s_x$ and $s_z$  
to uncover the entanglement structure of the dual many-body system.  

In the small $\bar{l}_{x,z}\ll 1$ regime (i.e., $r_c^{-1}\ll l_{x,z}\ll b$), the length dependence of $s_{x,z}$ is consistent with the pure AdS results in \cite{Ryu:2006ef}, independently of the value of $\bar{M}$. By fitting the numerical data, we find
\bea
\begin{split}
s_x&=-2\pi^{3/2}\left(\frac{\Gamma(\frac{2}{3})}{\Gamma(\frac{1}{6})}\right)^3l_x^{-2}+\tilde{s}_x(M, l_x, r_c)\,,\\
s_z&=-2\pi^{3/2}\left(\frac{\Gamma(\frac{2}{3})}{\Gamma(\frac{1}{6})}\right)^3l_z^{-2}+\tilde{s}_z(M, l_z, r_c)\,,
\end{split}
\eea
where we have extracted the leading divergent term in $l$ for $s_x$ and $s_z$.

Notice that in the NLSM phase, the bulk geometry is extremely anisotropic, i.e.,  
in the IR region we have ${\bf z}\simeq 10.908$. 
This suggests that we should use different widths of the strip 
$l_x$ and $l_z$ to efficiently probe the IR geometry. It turns out that, in order to well distinguish the three IR phases, we need to choose at least
$\bar{l}_x\sim O(10), \bar{l}_z\sim O(10^4)$. 

\begin{figure}[h!]
  \centering
\includegraphics[width=0.45\textwidth]{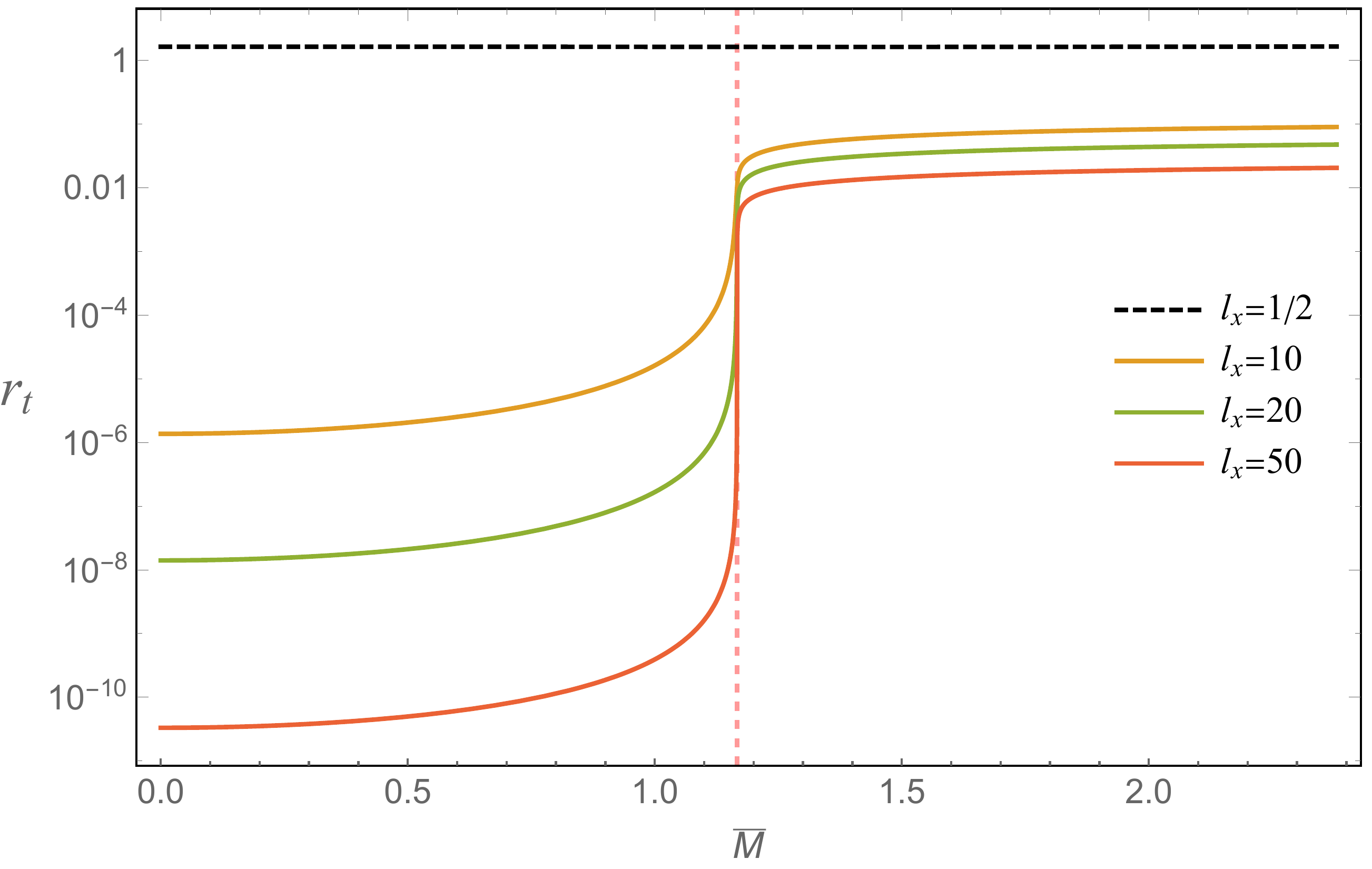}~~~
\includegraphics[width=0.45\textwidth]{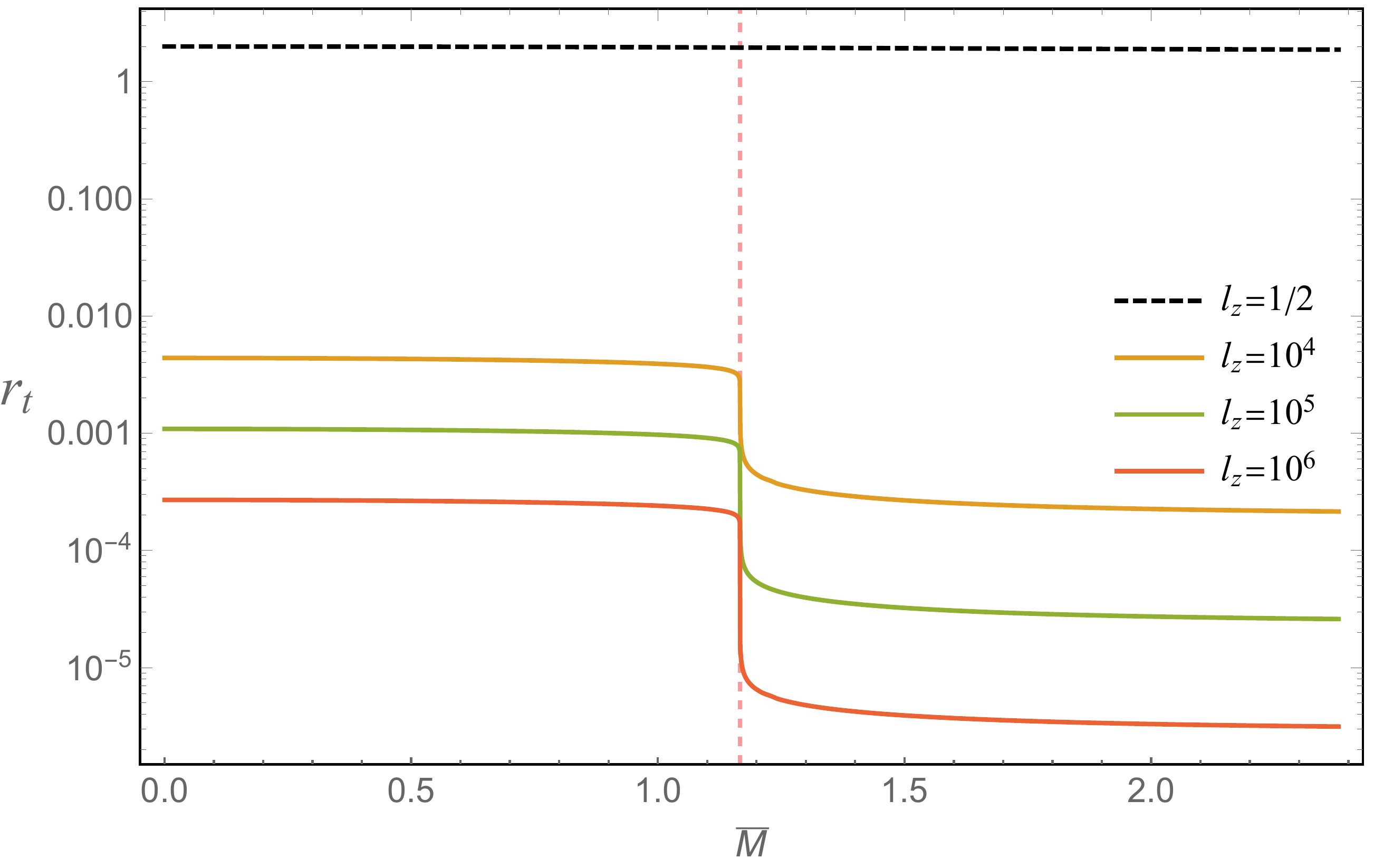}
  \caption{\small The location of the turning points $r_t$ across the topological quantum phase transition choosing $l$ along $x$-direction ({\bf left}) and $z$-direction ({\bf right}).}
  \label{fig:zstar}
\end{figure}

In Fig.\ref{fig:zstar}, we show the turning points $r_t$ as functions of $\bar{M}$ for different values of the strip width. For small $\bar{l}$ (dashed black curve), the location of the turning point is not able to distinguish the two phases, nor to locate the QCP, since the RT surface only probes the geometry near the UV boundary. On the contrary, for large  $\bar{l}_x\gtrsim 10$ or $\bar{l}_z \gtrsim 10^4$, the turning points $r_t$ increase/decrease (depending on the direction of the strip) drastically across the quantum phase transition. This behavior can be understood as follows. 
When $\bar{l}$ is large enough,  the RT surface probes the IR geometry, and the turning points with respect to $\bar{l}_x$ and $\bar{l}_z$ obey different power laws in the three phases, as summarized in Table. \ref{table:lx} and \ref{table:lz}. 
Due to the fact that ${\bf z}=10.908,2.968$ and $1$ for the NLSM phase, the QCP, and the trivial phase respectively, it is reasonable that the turning points show sharp changes during the phase transition. 

\begin{table}[h]
\centering
\setlength{\tabcolsep}{7mm}{
\begin{tabular}{|c|c|c|c|}
\hline
 &  UV & Intermediate & IR \\ \hline
$r_t$ & $l_x^{-1}$ & $l_x^{-2/\alpha_c}$ & $l_x^{-2/\alpha}$ \\ \hline
$s_x$ & $l_x^{-2}$ & $l_x^{-(1+{\bf z}_c)}$ & $l_x^{-(1+{\bf z})}$
\\ \hline
$c_x$ & $l_x^{0}$ & $l_x^{1-{\bf z}_c}$ & $l_x^{1-{\bf z}}$\\ \hline
\end{tabular}}
\caption{Summary of the power laws w.r.t $l_x$ in the three energy ranges for different phases. Notice that the parameters $\alpha,{\bf z}$ in the IR depend on the IR phase. The intermediate scalings appear only in the vicinity of the QCP, $\bar{M}\sim \bar{M}_c=1.1667$.}
\label{table:lx}
\end{table}

\begin{table}[h]
\centering
\setlength{\tabcolsep}{7mm}{
\begin{tabular}{|c|c|c|c|}
\hline
 &  UV & Intermediate & IR \\ \hline
$r_t$ & $l_z^{-1}$ & $l_z^{-2/\delta_c}$ & $l_z^{-2/\delta}$ \\ \hline
$s_z$ & $l_z^{-2}$ & $l_z^{-2/{\bf z}_c}$ & $l_z^{-2/{\bf z}}$
\\ \hline
$c_z$ & $l_z^{0}$ & $l_z^{2-2/{\bf z}_c}$ & $l_z^{2-2/{\bf z}}$\\ \hline
\end{tabular}}
\caption{Summary of the power laws w.r.t $l_z$ in the three energy ranges for different phases. Notice that the parameters $\alpha,{\bf z}$ in the IR depend on the IR phase.  The intermediate scalings appear only in the vicinity of the QCP, $\bar{M}\sim \bar{M}_c=1.1667$.}
\label{table:lz}
\end{table}

\begin{figure}[h!]
  \centering
\includegraphics[width=0.45\textwidth]{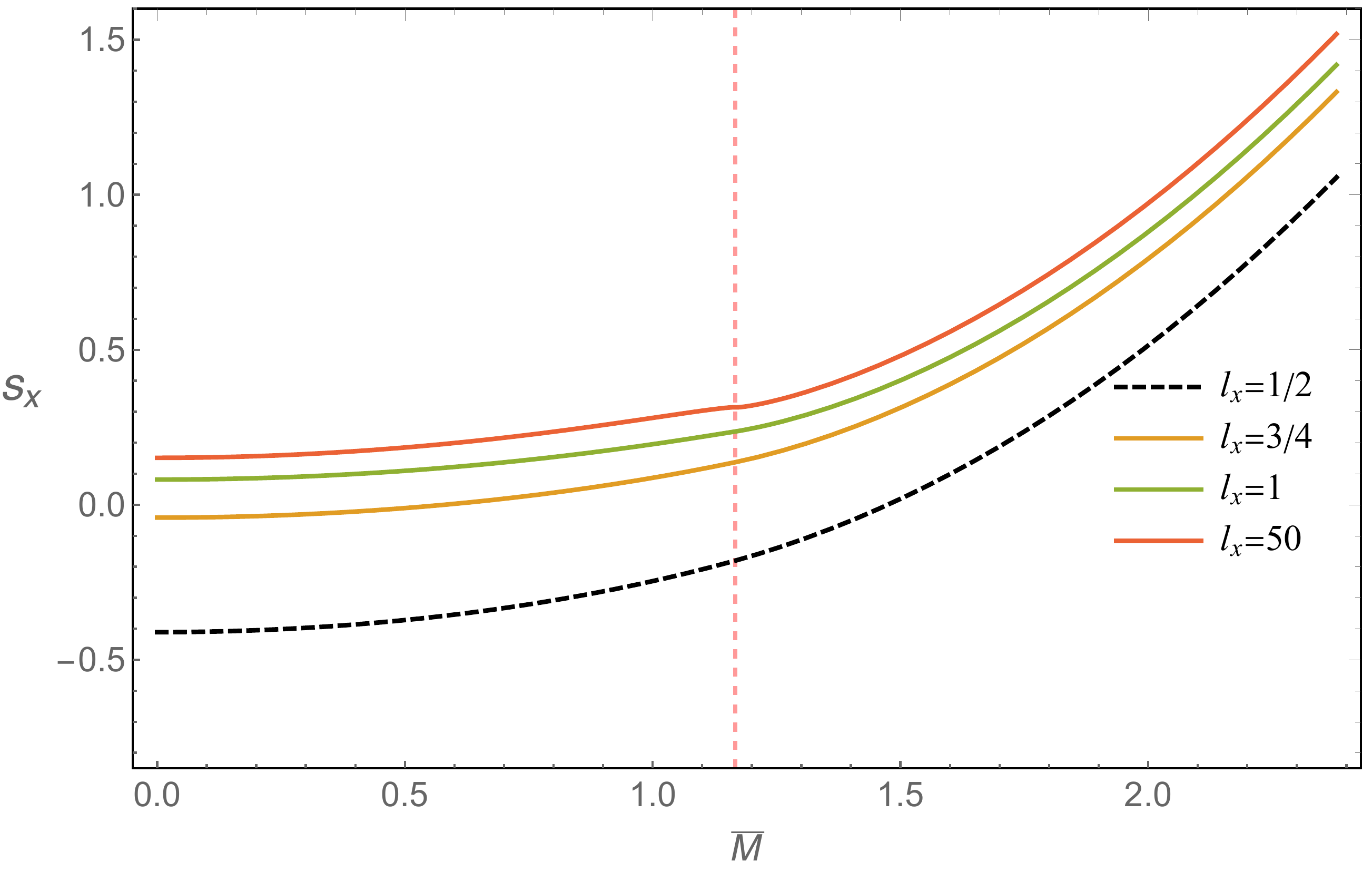}
\hspace{0.1pt}
\includegraphics[width=0.45\textwidth]{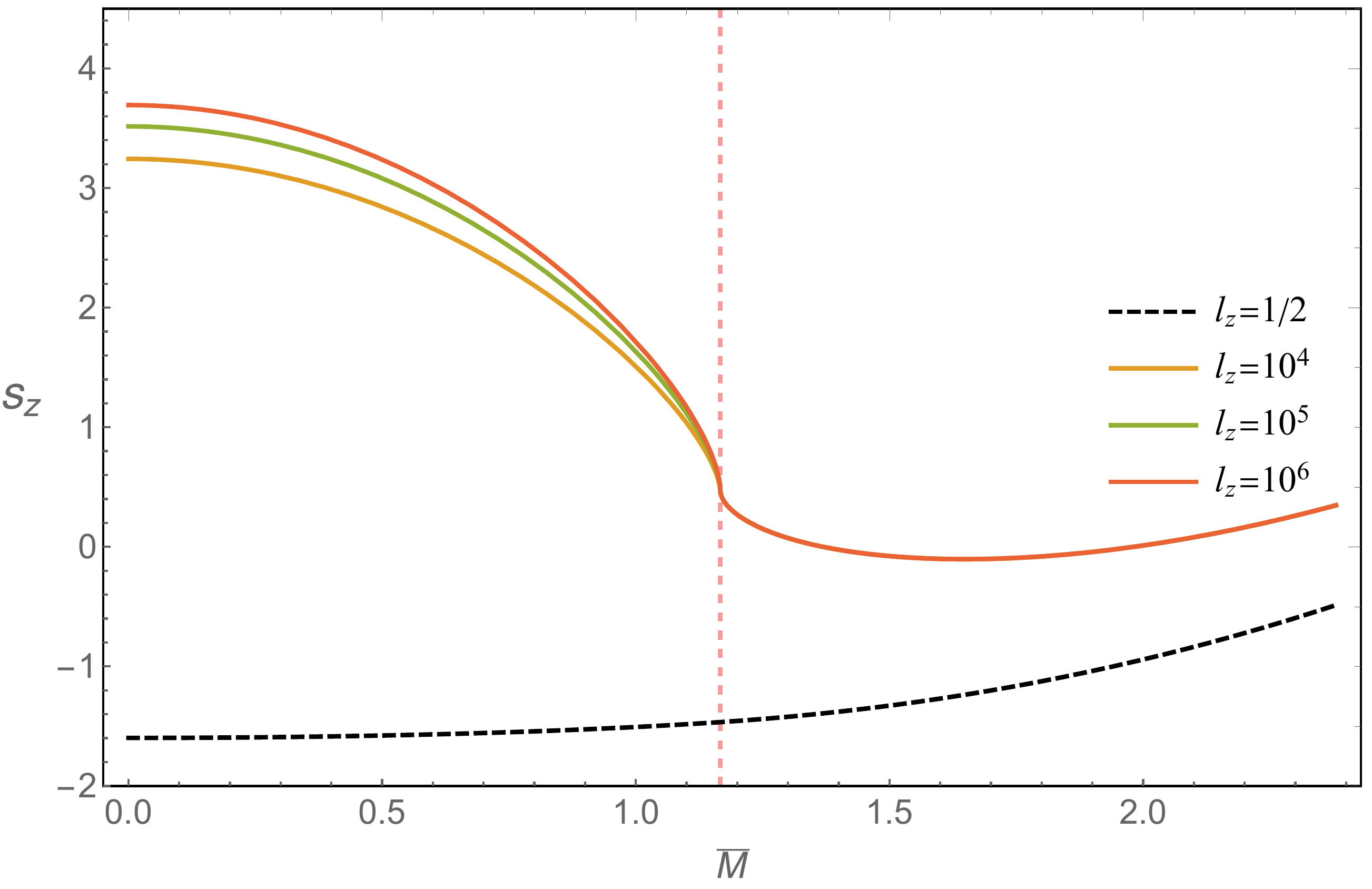}
  \caption{\small Entanglement entropies 
  $s_x$ ({\bf left}) and $s_z$ ({\bf right}) as a function of 
  $\bar{M}$ for different values of $l$ along the $x$ and $z$  directions respectively. The figures are done for $\bar{T}=0$. 
 }
  \label{fig:EE}
\end{figure}

In Fig.\ref{fig:EE}, we show the entanglement entropies $s_x, s_z$ as a function of $\bar{M}$ for $\bar{T}=0$. 
For small $l_i=1/2$ (dashed black curve), since the RT surface is not able to  probe the IR geometries, the EE $s_i$ are smooth and increase monotonically as $\bar{M}$ increases. In order words, for small values of $l_i$, the EE is not sensitive to the topological quantum phase transition. 
For larger values of $l_x$, a very mild feature starts to appear at the location of the quantum critical point, which can be detected by inspecting the derivative $\partial s_x/\partial \bar{M}$ (which will be shown in Fig.\ref{fig:EEslope} in Sec.\ref{locateQCP} ). The EE in the anisotropic direction $s_z$ shows a more pronounced (but qualitatively similar) feature at the quantum critical point, at which its derivative $\partial s_x/\partial \bar{M}$ diverges. 

At finite temperature, we can perform a similar analysis as in the zero temperature case, and study the behavior of the entanglement entropy. 
It is known that the RT surface cannot penetrate the horizon for static situations \cite{Hubeny:2012ry}.  When the width of the strip $l$ approaches infinity, the RT surface approaches closely to the horizon, $r_t \rightarrow r_0$. 
As a consequence, $c_z\rightarrow \frac{1}{2}l_z^3 f\sqrt{h}|_{r_0}$ as $l_z\rightarrow \infty$. 
Therefore, we can take the thermodynamic entropy density as an approximation for $c_z$ to see the behavior of the order parameter, Eq.\eqref{eq:orderM}, at finite temperature.

\begin{figure}[h!]
  \centering
\includegraphics[width=0.45\textwidth]{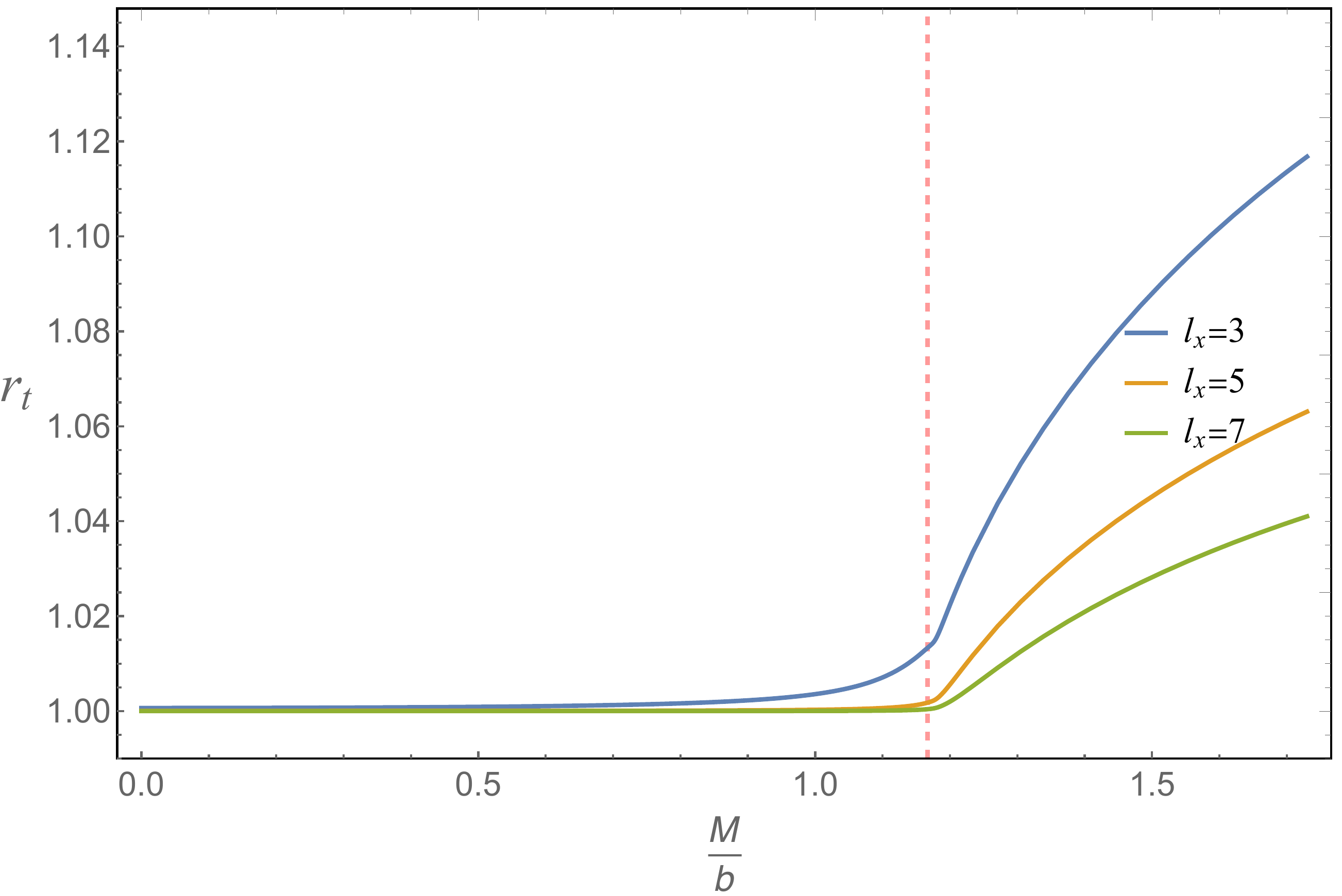}~~~
\hspace{0.1pt}
\includegraphics[width=0.45\textwidth]{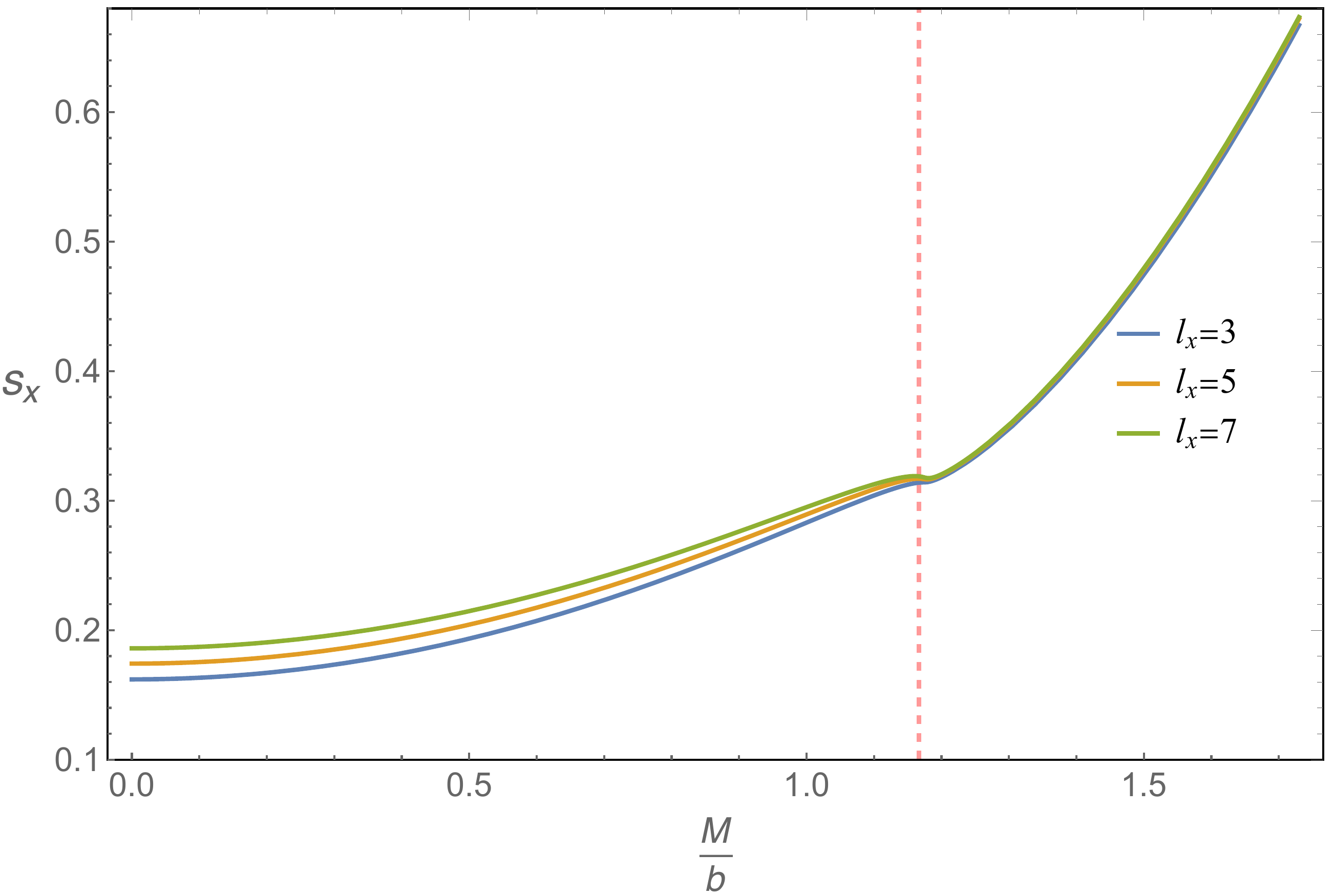}\\
\includegraphics[width=0.45\textwidth]{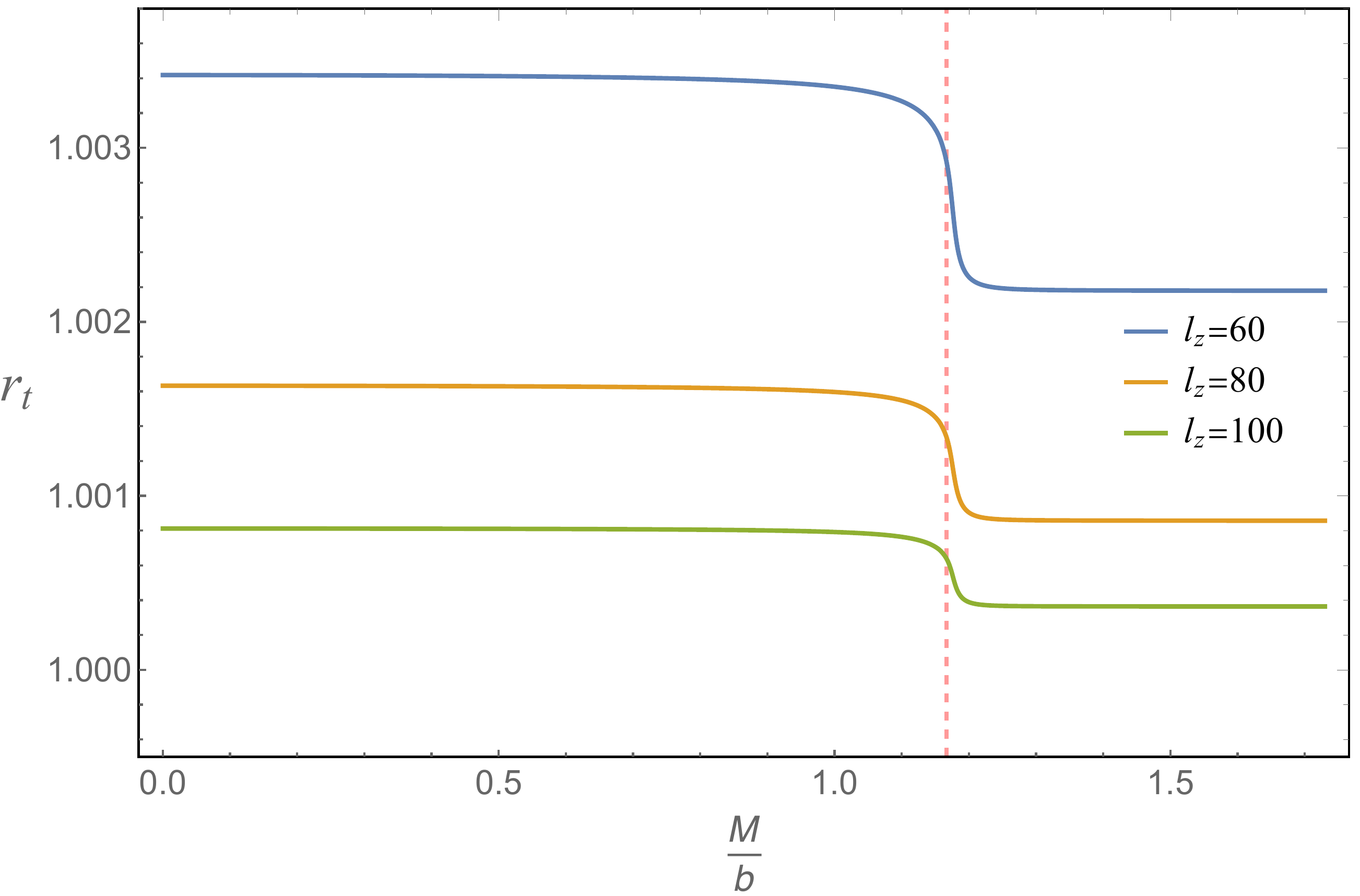}~~~
\hspace{0.1pt}
\includegraphics[width=0.45\textwidth]{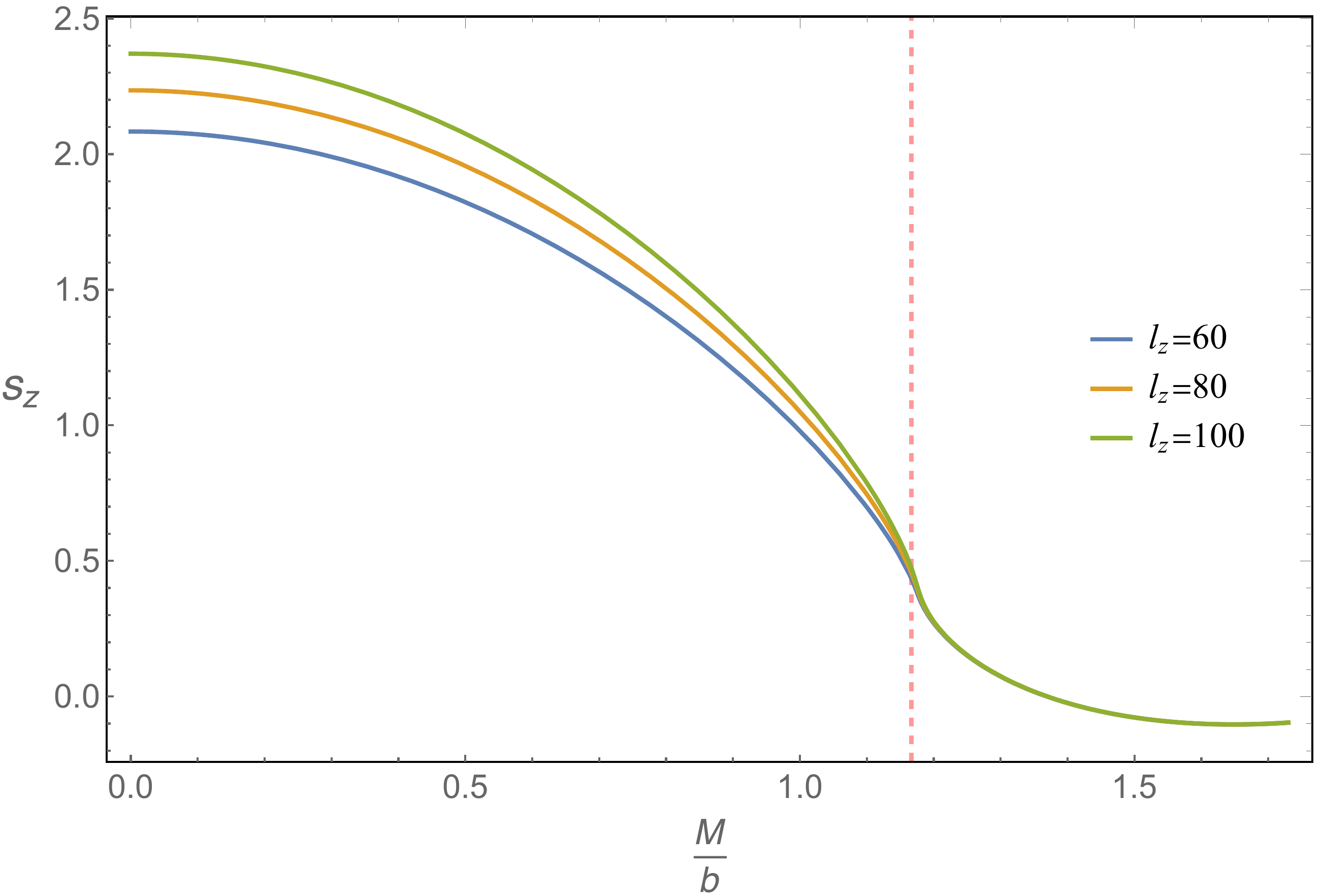}
  \caption{\small The turning point $r_t$ and the EE $S_i$ for the strip along $x$ ({\bf top panel}) and $z$ ({\bf bottom panel}) directions at low temperature $T/b=0.005$.}
  \label{fig:finiteTsz}
\end{figure}

The turning points $r_t$ and entanglement entropies $s$ as a function of $M/b$ for the strip geometry with the width along $x$ direction and $z$ direction are shown in Fig.\ref{fig:finiteTsz}. At low temperatures, $s_x$ and $s_z$ retain the zero temperature behaviors shown in Fig.\ref{fig:EE}, which then gradually disappear as the temperature increases. In other words, the fingerprints of the quantum critical point and the TQPT are still visible at low temperature, within the so-called quantum critical region. We will discuss this point furthermore in Sec.\ref{locateQCP}. 

\subsection{Order parameter for nodal topology}

In the previous subsection, we have computed the entanglement entropy $S_i$ for the strip geometries of width $l_i$ ($i=x,y,z$) and length $L\rightarrow \infty$ using the Ryu-Takayanagi (RT) prescription \cite{Ryu:2006bv}. In this subsection, motivated by the weakly coupled results in \cite{PhysRevB.95.235111}, we construct opportune physical quantities to characterize the topological phase transition and probe the nodal topology. 

For the entanglement entropy in these three configurations, due to the isotropy in the $x$-$y$ plane, we have $S_x=S_y$. 
We have already defined 
\be\label{eq:c-fun}
c_i=4\,G\frac{l_i^3}{L^2}\frac{\partial S_i}{\partial l_i}\,,
\ee
in \eqref{eq:defci} where $G$ is the Newton constant. These quantities are known as $c$-functions and can be used to parameterize the number of degrees of freedom along the renormalization group flow. Notice the similarity between \eqref{eq:c-fun} and the quantity defined in \eqref{wwk}. At zero temperature, \eqref{eq:c-fun} can be further simplified as $c_i=\frac{1}{2}\mathfrak{C}_il_i^3$
with $\mathfrak{C}_i= f\sqrt{u}{|}_{r_t}$. $r_t$ is the turning point of the extremal RT surface corresponding to the strip with width $l_i$. In the limit of $l_i \rightarrow \infty$, the turning point approaches the location of the IR horizon, $r=0$. For isotropic systems, all $c$-functions are equal and obey the so-called $c$-theorem \cite{Myers:2012ed,Liu:2013una}. For anisotropic systems, an analogous $c$-theorem has been proposed in \cite{Chu:2019uoh}, and utilized in the context of WSM in \cite{Baggioli:2020cld}. 

\begin{figure}[h!]
  \centering
\includegraphics[width=0.6\textwidth]{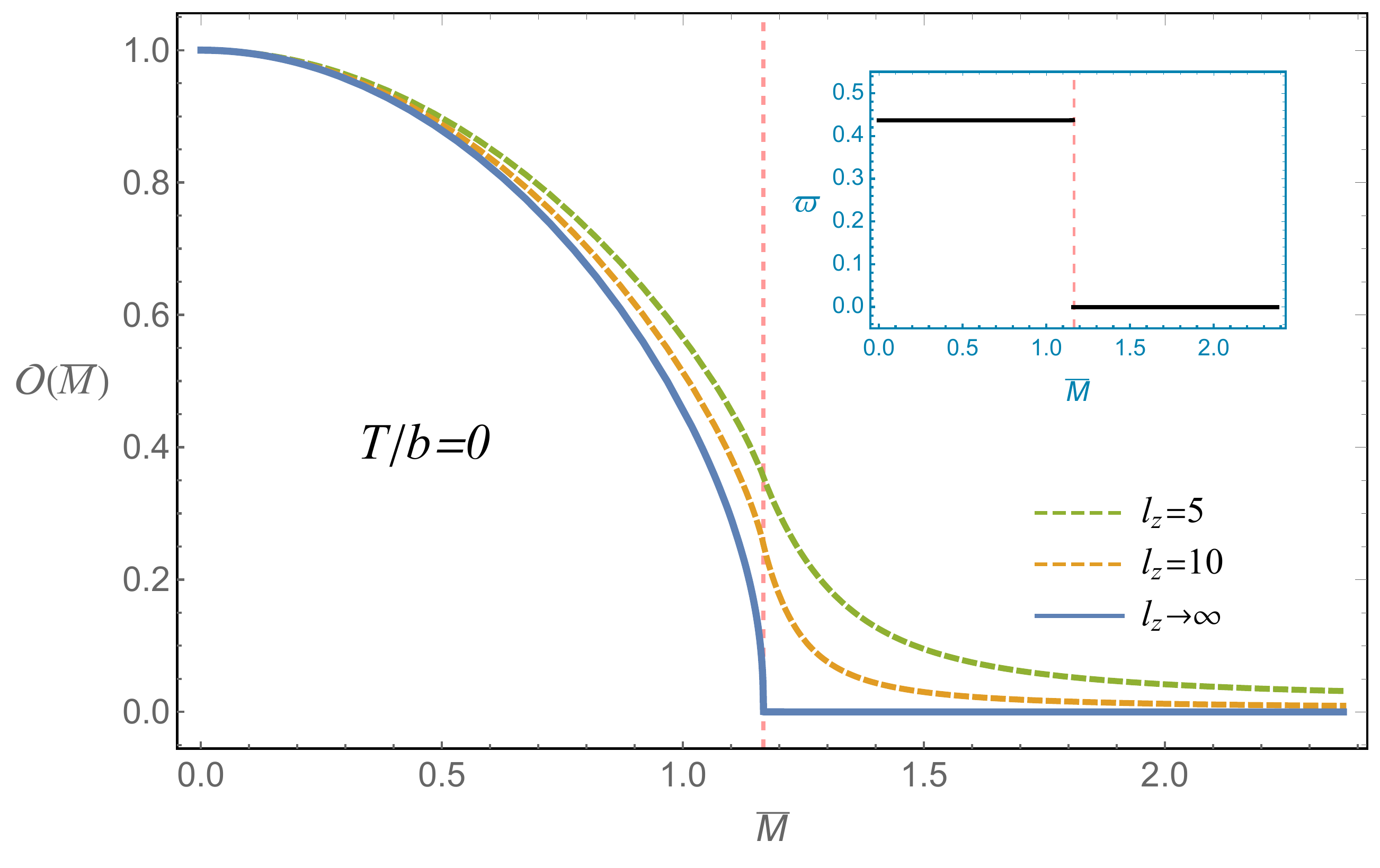}
  \caption{The order parameter $\mathcal{O}$, Eq.\eqref{eq:orderM}, as a function of $\bar{M}$ across the topological quantum phase transition. The normalized ratio $c_z(\bar{M})/c_z({\bar{M}=0})$ at $T=0$ converges to the order parameter (solid line) as $l_z$ increases. The inset shows the coefficient $\varpi(\textbf{z})$ defined in Eq.\eqref{deforder}.}
  \label{fig:order1}
\end{figure}

In order to characterize the quantum phase transition and probe the nodal topology, we define the following order parameter 
\bea
\label{eq:orderM}
\mathcal{O}(\bar{M})\equiv \lim_{l_z\to
\infty}\frac{c_z(\bar{M})}{c_z(\bar{M}=0)}\,,
\eea
where $\bar{M}=M/b$, and $z$ is the direction along the anisotropy. The  behavior of this quantity as a function of $\bar{M}$ at zero temperature is shown in Fig.\ref{fig:order1}. 

At zero temperature, $\mathcal{O}$ is zero in the topologically trivial phase and becomes non-zero in the topological NLSM phase. Its behavior across the quantum critical point is continuous and follows a power-law scaling $\mathcal{O}\propto (\bar{M}_c-\bar{M})^\xi$, with $\xi\approx 0.39$. This scaling exponent is different from the weakly-coupled theory  \eqref{eqL:imNLSM}, for which $\xi= 0.5$ (mean field behavior). Thermal effects and finite $l_i$ corrections modify the sharp transition into a smooth crossover. At $T=0$, we can analytically prove (see appendix \ref{appB}) that:
\begin{equation}\label{deforder}
    \mathcal{O}(\bar{M})= \lim_{l_z \rightarrow \infty}\varpi({\bf z}) \,l_z^{\frac{2}{{\bf z}}}\mathcal{B}_{xy}(r_t)\equiv \varpi({\bf z}) \beta_{xy}(\bar{M})\,.
\end{equation}
In this limit, the order parameter is independent of the geometry of the entanglement boundary region. $\varpi(\bf z)$ is a $\bar{M}$-independent function which is finite in the NLSM phase, zero in the trivial phase, and determined by the anisotropic IR exponent $\bf z$ (see inset in Fig.\ref{fig:order1}). The coefficient $\varpi({\bf z})$ in the topological phase can be computed analytically. Finally,
$\beta_{xy}(\bar{M})\propto 
r_t^{-\alpha} \mathcal{B}_{xy}(r_t)\,
$ in the IR limit $r_t\rightarrow 0$ which is equivalent to $l_z \rightarrow \infty$.
$\alpha$ is a parameter related to the scaling properties of the topological IR phase. The factor $l_z^{2/{\bf z}}$ in \eqref{deforder} cancels the $r_t$ factor in $\mathcal{B}_{xy}(r_t)$ and leads to a finite result.  Finally, $\beta_{xy}$ corresponds to the IR value for the field theory source $b_{xy}$.

Eq.\eqref{deforder} shares strong similarities with the weakly-coupled result, \eqref{eq:EEunique} or \eqref{wwk}. Nevertheless, it is important to emphasize a crucial difference between the two approaches. The weakly coupled field theory result is obtained at finite $l$. On the contrary, our proposed order parameter is defined only in the limit $l_z \rightarrow \infty$. From the holographic perspective, such a limit is necessary to push the RT surface deep into the bulk and therefore probe the IR geometries and the corresponding IR physics. Also, in the weakly coupled picture, only the gapless modes around the Fermi nodal line are considered while the contributions in EE from gapped modes are ignored. This can be thought as an IR limit of the theory, which in holography is realized as explained above.

In any case, despite we are not able to provide a formal proof, there is good evidence that the IR parameter $\beta_{xy}$ is proportional to the nodal line length $k_F$, as in the weakly-coupled picture. This can be seen from the fact that $\mathcal{B}_{xy}=0$ in the IR corresponds to the trivial phase, with no nodal line in the fermionic spectral function \cite{Liu:2018bye,Liu:2020ymx}. Additionally, $\mathcal{B}_{xy}\neq0$ implies the breaking of time reversal and charge conjugate symmetry, which is a distinctive feature of the NLSM phase with $k_F\neq0$.

The different scaling of $S$, and consequently of $c_i$, with respect to the length-scales $l_i$ is due to the anisotropic nature of the IR fixed point in the strongly-coupled holographic model. In the weakly-coupled case \cite{PhysRevB.95.235111}, in order to obtain the universal relation between the EE and the nodal line length, it is imperative to sum over the different directions. In our case, because of ${\bf z}_{\text{nlsm}}>{\bf z}_{\text{trival}}=1$, the EE related to the strip oriented along the anisotropic direction represents always the dominant contribution in the large $l$ limit. Because of this reason, $\sum_i c_i \approx c_z$, and our definition in \eqref{eq:orderM} is equivalent to that in \cite{PhysRevB.95.235111}. 

\begin{figure}[h!]
  \centering
\includegraphics[width=0.55\textwidth]{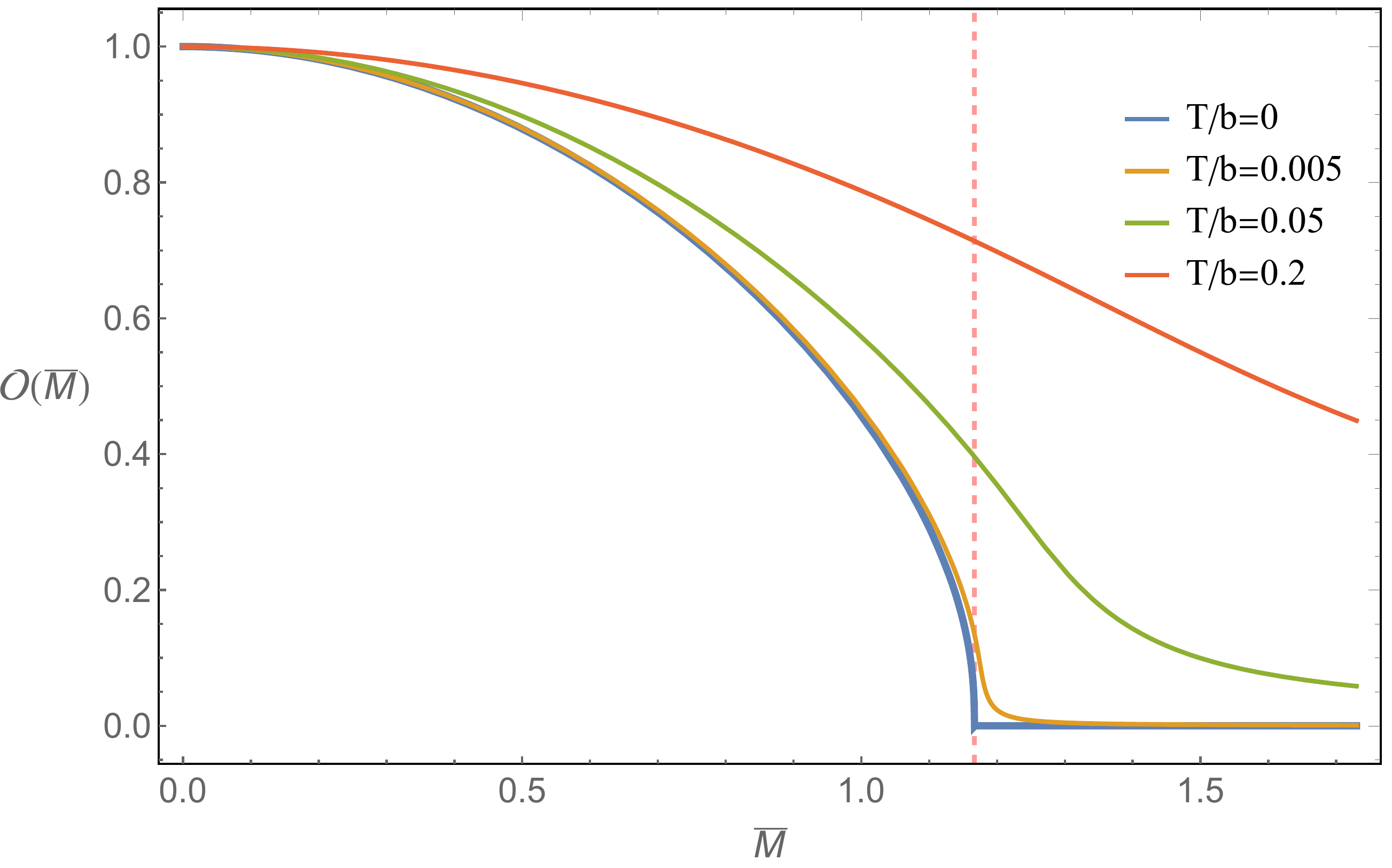}
  \caption{\small 
   The order parameter defined in \eqref{eq:orderM}, as a function of the external parameter $\bar{M}$ for different values of the dimensionless temperature.}
  \label{fig:finiteTOP}
\end{figure}

In order to explore the finite temperature region, we further plot the order parameter defined in \eqref{eq:orderM} at finite temperature in Fig.\ref{fig:finiteTOP}. As expected, thermal effects smooth out the sharp behavior at the critical point. Nevertheless, for low enough temperature, the order parameter is still able to identify a smooth crossover between the two different phases. Our results suggest the utility of this probe as an order parameter to characterize the phase transition also at small and finite temperature and to draw the whole quantum critical region. A more detailed analysis is needed to confirm our expectations at finite temperature, away from the quantum critical point. 

\subsection{Locating the quantum critical point}
\label{locateQCP}

In order to probe the quantum critical point further, we define a second quantity which is given by $\partial s_i/\partial \bar{M}$. Here, $s_i$ is the renormalized EE once the UV divergent terms in the EE $S_i$ are removed, i.e., $s_i$ does not depend on the UV cutoff anymore. When $l_i$ is large enough, the RT surfaces probe the IR geometries, $s_i$ is therefore sensitive to the IR properties of our many-body system, which carry the fingerprints of the TQPT.

Conceptually, this definition shares many similarities with the proposal of \cite{Osterloh:2022}. There, it was shown that the derivative of the ``entanglement of formation'' $C$ with respect to the external coupling $\lambda$ shows a sharp dip at a quantum critical point, signaling the divergence of non-local correlations in the critical region. Here, we run a similar argument in terms of the renormalized EE.

\begin{figure}[h!]
  \centering
\includegraphics[width=0.45\textwidth]{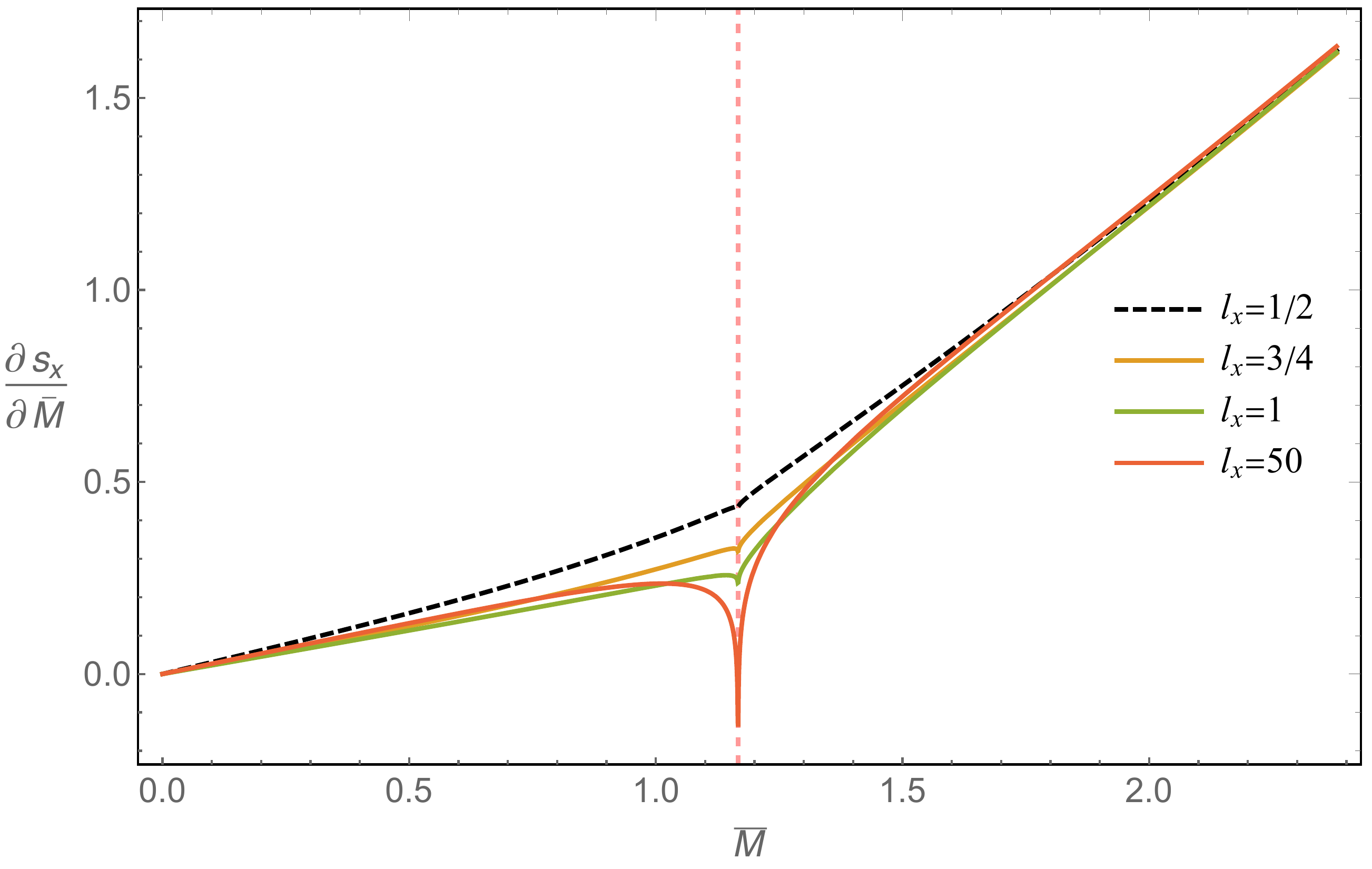} \includegraphics[width=0.45\textwidth]{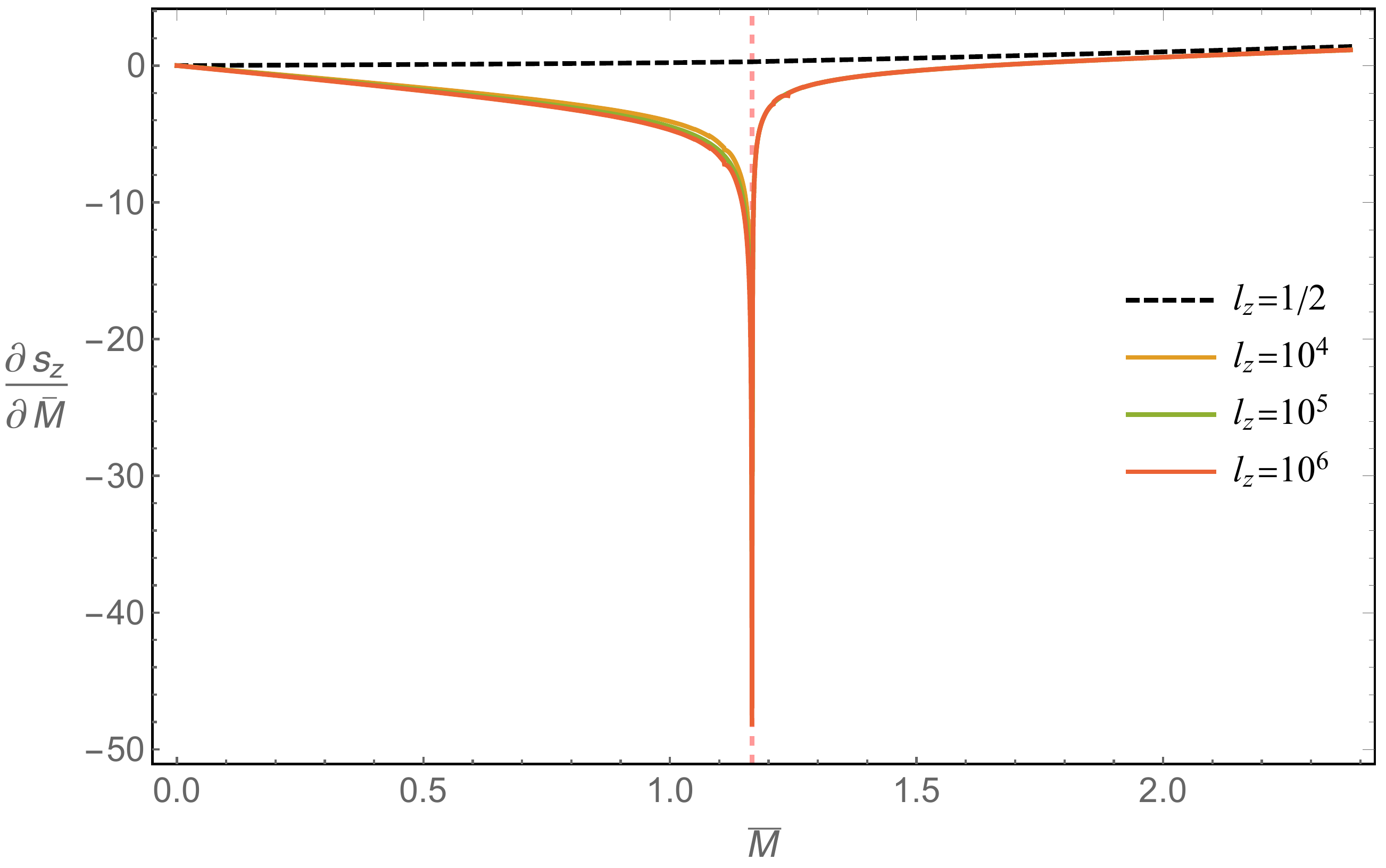}
  \caption{The derivative of the renormalized EE $s_i$ with respect to $\bar{M}$ for $l$ along the $x$ ({\bf top}) and $z$ ({\bf bottom}) directions.
  The vertical dashed line indicates the location of the QCP.}
  \label{fig:EEslope}
\end{figure}

The results at zero temperature are shown in Fig.\ref{fig:EEslope}. For large enough entanglement regions, $l_i \gg 1$, both derivatives in the parallel (with respect to anisotropy) and perpendicular directions display a clear signature at the QCP. In the limit of $l_i \rightarrow \infty$, in which the EE surface reaches the IR horizon of the geometry, the derivatives become divergent at the QCP. In analogy to the behavior of the correlation length in classical thermal phase transition, it is tempting to associate this feature to the divergence of non-local quantum correlations at the QCP. 

\begin{figure}[h!]
  \centering
\includegraphics[width=0.45\textwidth]{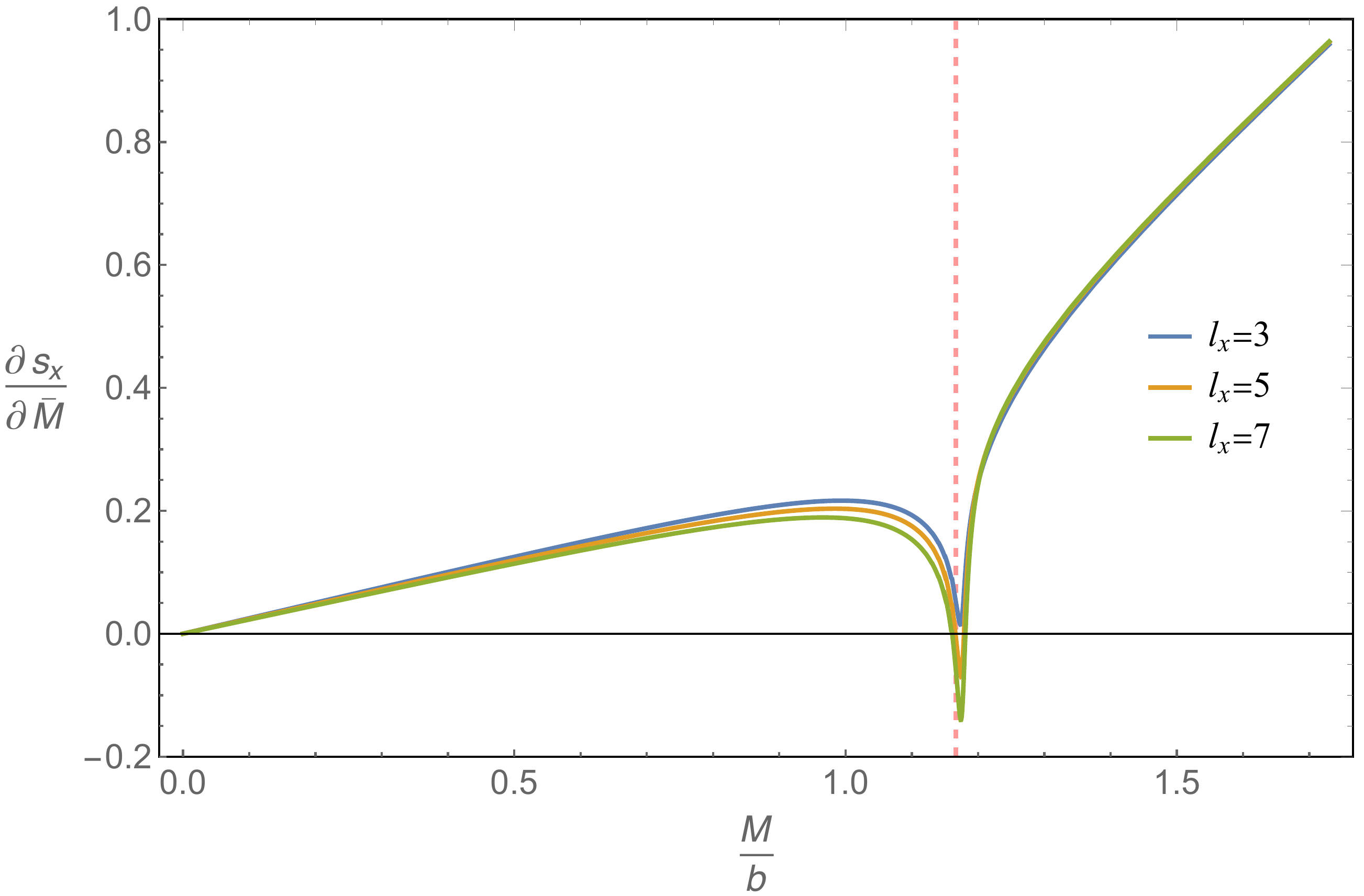}
\hspace{0.1pt}
\includegraphics[width=0.45\textwidth]{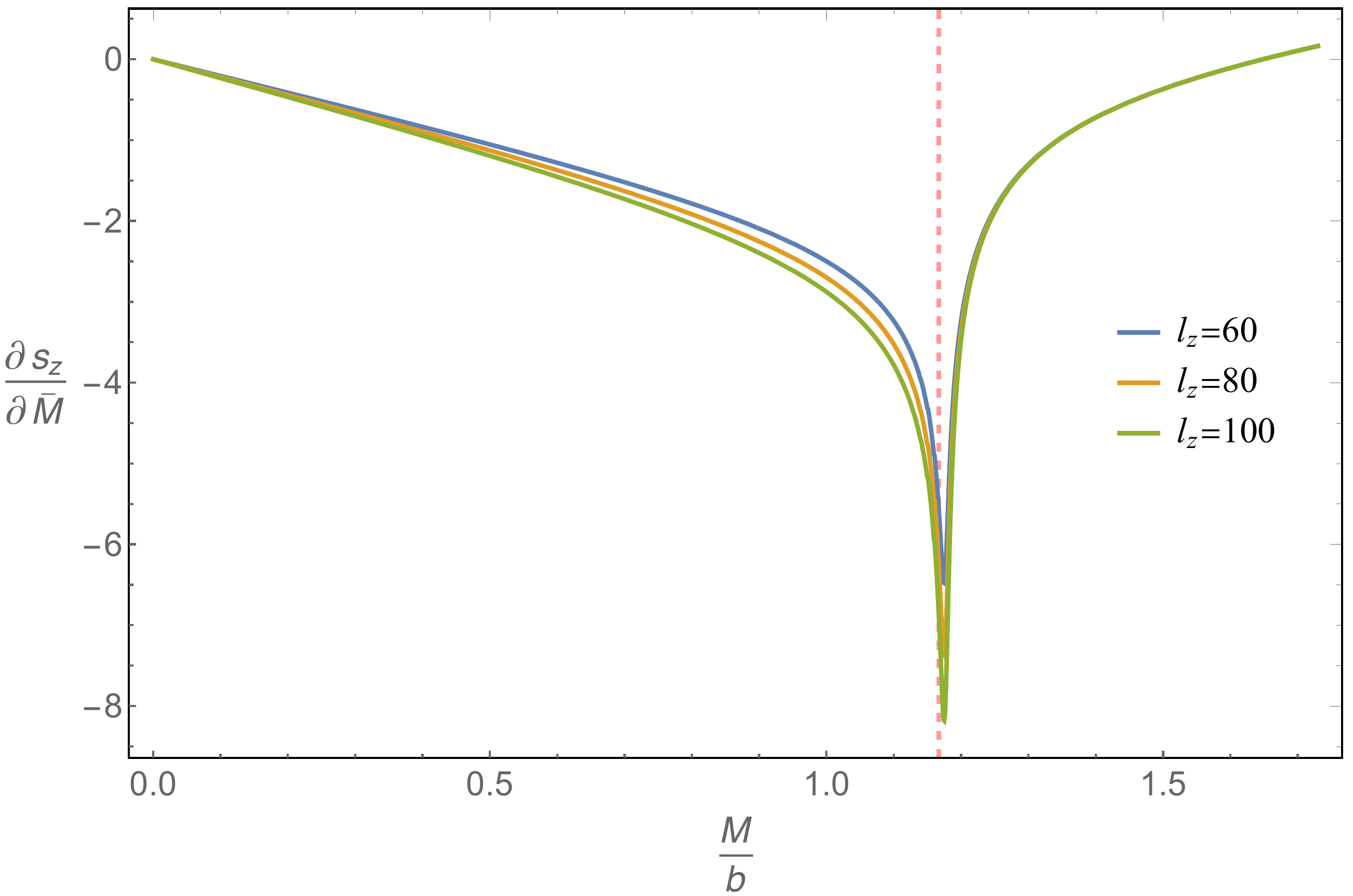}
  \caption{\small $\partial s_x/\partial \bar{M}$ and $\partial s_z/\partial \bar{M}$ at low temperature $T/b=0.005$ as a function of $\bar{M}$.
 Close to the QCP, both curves show a dip behavior similar to the case of zero temperature, Fig.\ref{fig:EEslope}.}
  \label{fig:finiteTpspM}
\end{figure}

\begin{figure}[h!]
  \centering
\includegraphics[width=0.55\textwidth]{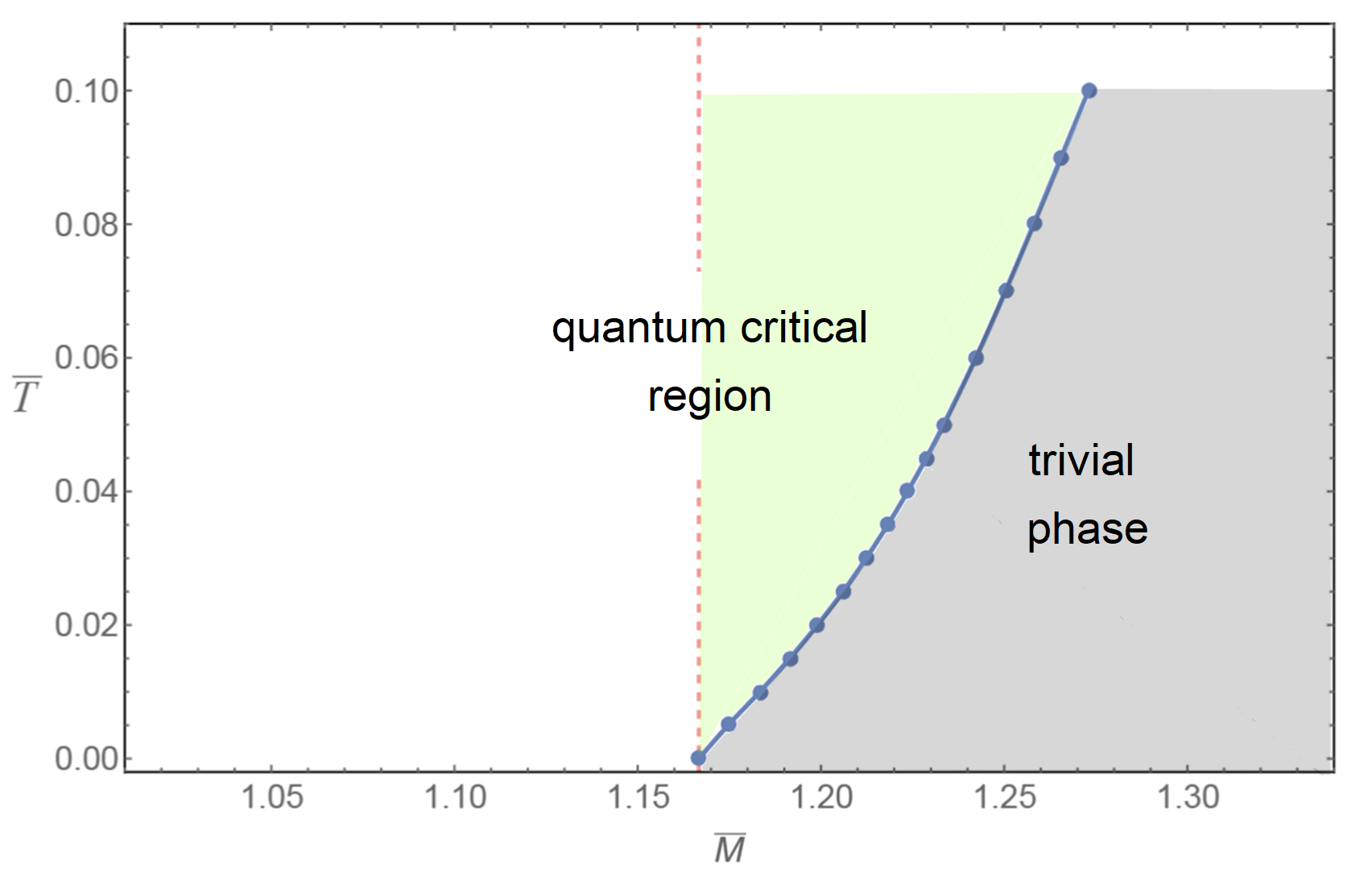}
  \caption{\small 
   By tracking the location of the minimum in $\partial s_z/\partial \bar{M}$ as a function of $T/b$, which resembles the edge of a quantum critical region between the topological and trivial phases. Further analysis, e.g., a comparison to the different transport properties of the two phases, is needed to verify the validity of this proposal.}
  \label{fig:QCR}
\end{figure}

In Fig.\ref{fig:finiteTpspM}, we plot the derivative of the renormalized EE $s_i$ as a function of the control parameter $\bar{M}$ at small and finite temperature. Thermal effects moves the minimum slightly away from the QCP. This indicates that our quantum information inspired quantities might be good probes not only for the QCP but for the quantum critical region as well. We can think of this feature as the onset of the quantum critical region and the fact that at finite temperature the transition between the two phases moving to the location of the zero temperature quantum critical point. The location of the minimum in the $\partial s_i/\partial \bar{M}$ derivatives is plotted as a function of the renormalized temperature in Fig.\ref{fig:QCR}. We conjecture that this minimum could locate the crossover between the two phases at finite temperature, and define therefore a corresponding quantum critical region. Importantly, this phenomenological criterion seems to be able to estimate only half of the quantum critical region. In order to confirm whether this is still a good criterion at finite temperature, further analysis is needed. It would be nice to perform a study of the transport properties and related scalings similar to that in \cite{Rodgers:2021azg}, to check this further and see whether the quantum critical fan defined with transport coincides with our line in Fig.\ref{fig:QCR}. This would confirm or disprove the validity of the proposed criterion to draw a possible quantum critical region.

\section{Discussion}
\label{Sec:conclusion}
Using a holographic model for strongly-coupled nodal line semimetals, we show that quantum information quantities related to entanglement entropy are efficient probes for nodal topology and for the topological quantum phase transition between NLSM and topologically  trivial phases. Inspired by field theory studies \cite{PhysRevB.95.235111}, we propose an order parameter which displays a critical and beyond mean-field behavior across the topological phase transition. Moreover, in analogy with the findings of \cite{Osterloh:2022}, we find that the derivative of the renormalized EE with respect to the external coupling driving the quantum phase transition diverges at the critical point, signaling the explosion of non-local quantum correlations. 

Interestingly, our findings are robust against thermal effects and indicate that quantum information observables might be functional to outline and describe quantum critical regions in many-body systems. In other words, the full structure of the renormalization group flow from the UV to the IR fixed points might be retrieved in the scaling behavior of those quantities, providing useful insights away from the quantum critical point.

Finally, we comment on further observations and open questions. 
The first interesting observation is that, for the specific case of ${\bf z}=2$ (which can be achieved holographically by using a specific value for
$\lambda_2$), in the limit of large $l_z$, the dependence of the entanglement entropy $s_z$ with respect to $l_z$ is inversely proportional to $l_z$, as shown in table \ref{table:lz}.
This coincides with the result in Eq. \eqref{pp} obtained from field theory. The value of ${\bf z}=2$ is also unique with respect to the longitudinal DC conductivity $\sigma^{\text{DC}}_{zz}$ in topological NLSM phase, computed in \cite{Rodgers:2021azg}. Indeed, that is the only value of ${\bf z}$ for which $\sigma^{\text{DC}}_{zz}$ in the zero temperature limit remains finite. It is therefore interesting to pursue further study to explore the interplay between the entanglement entropy and transport, in particular for the ${\bf z}=2$ scenario. 

On the other hand, Ref.\cite{huh2016long} found that in Dirac NLSM with strong long-range Coulomb interactions, and in the large $N$ limit, the DC conductivities are proportional to radius of the nodal ring $k_F$. It would be interesting to study if any holographic model could realize such a relation in Weyl NLSM, and if the latter is ultimately universal or not.

In addition, differently from the field theory framework, accurately locating the radius of the nodal Fermi surface in holography is a non trivial task since the fermions can only be introduced as a probe (as done in \cite{Liu:2020ymx}), and the positions of the poles in the fermionic spectral functions can be strongly dependent on the probe action chosen. 
In this direction, semi-holography might be a good candidate to simulate the dynamics of the fermionic excitations coupled to a strongly interacting sector state which determines the properties of their self-energy \cite{Faulkner:2010tq}. We leave this study for the future.

\acknowledgments

We thank Karl Landsteiner and Ya-Wen Sun for useful comments on a preliminary draft of this work.
M.B. and X.-M.W. acknowledge the support of the Shanghai Municipal Science and Technology Major Project (Grant No.2019SHZDZX01). M.B. acknowledges the sponsorship from the Yangyang Development Fund. Y.L. is supported by the National Natural Science Foundation of China grant No.11875083.

\appendix
\section{Details about the holographic solutions}
\label{appA}
The first holographic model for strongly coupled NLSM has been proposed in \cite{Liu:2018bye}. In the main text, we consider an improved version of it which was introduced in \cite{Liu:2020ymx}. The main difference is given by a Chern-Simons term for the complex two-form field, as opposed to a canonical kinetic term. The role of such a term is to impose the self-duality constraint on the dual field theory two-form operators (see \cite{Liu:2020ymx} for details). For convenience, we report the complete gravitational action, corresponding to \eqref{eq:action} in the main text, here:
\bea
\begin{split}
S&=\int d^5x\sqrt{-g}\,\bigg[R+12-\frac{1}{4}\mathcal{F}^2-\frac{1}{4}F^2
+\frac{\alpha}{3}\epsilon^{abcde}A_a \Big(3\mathcal{F}_{bc}\mathcal{F}_{de}+F_{bc}F_{de}\Big)
-(D_a \Phi)^*(D^a\Phi)\\&
-V_\Phi
-\frac{i}{6\eta}\epsilon^{abcde} \Big(B_{ab}H_{cde}^*-B_{ab}^* H_{cde}\Big)
-V_B-
\lambda|\Phi|^2B_{ab}^*B^{ab}\bigg]\,,
\label{eq:action2}
\end{split}
\eea
where $D_a\Phi=\nabla_a\Phi-i q_1 A_a\Phi$. 
The potentials are chosen to be 
$
V_\Phi=m_1^2 |\Phi|^2+\frac{\lambda_1}{2} |\Phi|^4$ and 
$
V_B=m_2^2 B^*_{ab}B^{ab}+\frac{\lambda_2}{2}(B^*_{ab}B^{ab})^2\,
$.

The equations of motion following from the gravitational action \eqref{eq:action2},
are given by
\bea
\label{eq:ein1}
\begin{split}
0&=R_{ab}-\frac{1}{2}g_{ab}(R+12)-T_{ab}\,,\\
0&=\nabla_b \mathcal{F}^{ba}+2\alpha \epsilon^{abcde} F_{bc}\mathcal{F}_{de}\,,\\
0&=\nabla_b F^{ba}+\alpha \epsilon^{abcde} (F_{bc}F_{de}+\mathcal{F}_{bc}\mathcal{F}_{de})
-iq_1\big(\Phi^*D^a\Phi-(D^a\Phi)^*\Phi\big)
+\frac{q_2}{\eta}\epsilon^{abcde} B_{bc}B_{de}^*\,,\\
0&=D_a D^a\Phi-\partial_{\Phi^*} V_\Phi-\lambda\Phi B_{ab}^*B^{ab}\,,\\
0&=\frac{i}{3\eta} \epsilon_{abcde} H^{cde}-m_2^2 B_{ab}-\lambda_2B^*_{cd}B^{cd}B_{ab}-\lambda \Phi^*\Phi B_{ab}\,,
\end{split}
\eea
where 
\bea
\begin{split}
T_{ab}=&\,\frac{1}{2}\Big[\mathcal{F}_{ac}\mathcal{F}_{b}^{~c}-\frac{1}{4}g_{ab}\mathcal{F}^2\Big]+
\frac{1}{2}\Big[F_{ac} F_{b}^{~c}-\frac{1}{4}g_{ab}F^2\Big]+\frac{1}{2}\big((D_a\Phi)^*D_b\Phi+(D_b\Phi)^*D_a\Phi\big)\\
&+(m_2^2+\lambda_2B^*_{cd}B^{cd}+\lambda|\Phi|^2)(B_{ac}^*B_b^{~c}+B_{bc}^*B_a^{~c})\\
&
-\frac{1}{2}\Big((D_c \Phi)^*(D^c\Phi)+V_\Phi+V_B+\lambda|\Phi|^2B_{cd}^*B^{cd} \Big)g_{ab}\,
\end{split}
\eea
is the energy-momentum tensor.

The ansatz for the zero temperature solutions reads
\bea
\label{eq:ztansatz2}
\begin{split}
&ds^2=\frac{dr^2}{r^2}+u(-dt^2+dz^2)+f(dx^2+dy^2)\,,\\
&\Phi=\phi\,,\,B_{xy}=-B_{yx}=\mathcal{B}_{xy}\,,\,B_{tz}=-B_{zt}=i\mathcal{B}_{tz}\,.
\end{split}
\eea
The corresponding equations of motion are given by
\bea
\begin{split}
0&=\frac{u''}{u}+\frac{f''}{f}-\frac{1}{3}\left(\frac{u'}{u}-\frac{f'}{f}\right)^2+\frac{u'}{ru}+\frac{f'}{rf}-\frac{8}{r^2}+\frac{2}{3}\phi'^2
+\frac{2}{3r^2}\left(m_1^2\phi^2+\frac{\lambda_1}{2}\phi^4-\frac{\lambda_2}{2}B^4\right)\,,\\
0&=\frac{u''}{u}-\frac{f''}{f}+\frac{1}{r}\left(\frac{u'}{u}-\frac{f'}{f}\right)-4(m_2^2+\lambda\phi^2+\lambda_2B^2)\left(\frac{\mathcal{B}_{tz}^2}{r^2u^2}+\frac{\mathcal{B}_{xy}^2}{r^2f^2}  \right)\,,\\
0&=\phi''+\left(\frac{u'}{u}+\frac{f'}{f}+\frac{1}{r}\right)\phi'-\left(m_1^2+\lambda_1\phi^2+\lambda B^2\right)\frac{\phi}{r^2}\,,\\
0&=\mathcal{B}_{tz}'-\frac{\eta u}{2rf}(m_2^2+\lambda \phi^2+\lambda_2B^2)\mathcal{B}_{xy}\,,\\
0&=\mathcal{B}_{xy}'-\frac{\eta f}{2ru}(m_2^2+\lambda \phi^2+\lambda_2B^2)\mathcal{B}_{tz}\,,
\end{split}
\eea
where $B^2\equiv B^*_{ab}B^{ab}$.

The asymptotic behaviors of the bulk fields at the UV boundary ($r\rightarrow \infty$) are
\be
\label{eq:uv}
\begin{split}
u\big{|}_{r\rightarrow \infty} &=r^2-b^2-\frac{M^2}{6}+\frac{1}{24}(8b^4+(2+3\lambda_1)M^4)\frac{\text{ln}(r)}{r^2}+\frac{u_2}{r^2}+...\,,\\
f\big{|}_{r\rightarrow \infty} &=r^2+b^2-\frac{M^2}{6}+\frac{1}{24}(8b^4+(2+3\lambda_1)M^4)\frac{\text{ln}(r)}{r^2}+\frac{f_2}{r^2}+...\,,\\
\phi\big{|}_{r\rightarrow \infty} &=\frac{M}{r}-\frac{M^3}{6}(2+3\lambda_1)\frac{\text{ln}(r)}{r^3}+\frac{\phi_2}{r^3}+...\,,\\
\mathcal{B}_{tz}\big{|}_{r\rightarrow \infty} &=br-2b^3\,\frac{\text{ln}(r)}{r}+\frac{b_{tz2}}{r}+...\,,\\
\mathcal{B}_{xy}\big{|}_{r\rightarrow \infty} &=br+2b^3\,\frac{\text{ln}(r)}{r}+\frac{b_{xy2}}{r}+...\,,\\
\end{split}
\ee
where the parameters satisfy
\be
\begin{split}
u_2+f_2&=\frac{2b^4}{3}+\frac{b\,b_{xy2}}{3}+\frac{M^4}{18}+\frac{\lambda}{6}b^2M^2+\frac{\lambda_1M^4}{24}-\frac{M\phi_2}{2}\,,\\
b_{tz2}+b_{xy2}&=-\lambda b M^2\,.
\end{split}
\ee
The geometry is asymptotically AdS$_5$ in the UV:
\bea
\lim_{r\rightarrow \infty}ds^2=\frac{dr^2}{r^2}+r^2(-dt^2+dz^2+dx^2+dy^2)\,.
\eea

The zero-temperature solutions present three independent scaling symmetries:
\bea
 \begin{split}
  \textbf{(I)}\quad   \{r^{-1}, t, x, y, z\}&\rightarrow \{\tilde{r}^{-1}, \tilde{t}, \tilde{x}, \tilde{y}, \tilde{z}\}\equiv c\{r^{-1}, t, x, y, z\}\,,\\ \{u, f, B_{\mu\nu}\} &\rightarrow \{\tilde{u}, \tilde{f}, \tilde{B}_{\mu\nu}\}\equiv c^{-2}\{u, f, B_{\mu\nu}\}\,,
  \end{split}
\eea
under which the metric $ds^2$ and the two form $B=B_{\mu\nu}dx^{\mu}dx^{\nu}$ remain unchanged. This symmetry can be used to fix $b=1$.
\begin{equation}
    \textbf{(II)}\quad ~~\{x, y\}\rightarrow \{\tilde{x}, \tilde{y}\}\equiv c\{x, y\}\,,\quad \{f, B_{xy}\} \rightarrow \{\tilde{f}, \tilde{B}_{xy}\}\equiv c^{-2}\{f, B_{xy}\}\,.\quad
\end{equation}

This symmetry allows us to fix the asymptotic behavior of $f(r)$ and to set one of the shooting parameters in the IR expansion.
\begin{equation}
    \textbf{(III)}\quad ~~~\{t,z\} \rightarrow \{\tilde{t},\tilde{z}\}\equiv c\{t,z\}\,,\quad \{u, B_{tz}\} \rightarrow \{\tilde{u}, \tilde{B}_{tz}\}\equiv c^{-2}\{u, B_{tz}\}\,.
\end{equation}
This symmetry allows us to set the UV asymptotics of $u(r)$ and to fix one of the shooting parameters in the IR expansion. We use the above symmetries to set $b=1$. 

\section{Additional results}
\label{appB}
We first show in appendix.\ref{app:B1} that at the large entangled region, the computations are simplified and we obtain a clear analysis on scaling behaviors. Then in appendix.\ref{app:B2} we further analytically prove the relation between entanglement and the nodal line length. 
\subsection{A quantum critical region and the holographic renormalization group flow}\label{app:B1}
When the entangled region becomes extremely large, the RT surface probes the IR geometry and the entanglement entropy reflects the scaling symmetries of the corresponding IR phase with respect to $l$. 

Eq.\eqref{eq:eqrt} can be re-written in the following form, 
\be\label{eq:ellx}
l_x=2\int_{r_t}^{r_{\text{cutoff}}} dr \frac{\mathfrak{C}_1}{\sqrt{g^{rr}g_{xx}(g_{xx}g_{yy}g_{zz}-\mathfrak{C}_1^2)}}+2\int_{r_{\text{cutoff}}}^\infty dr \frac{\mathfrak{C}_1}{\sqrt{g^{rr}g_{xx}(g_{xx}g_{yy}g_{zz}-\mathfrak{C}_1^2)}}\,
\ee
where $<r_0<r_{\text{cutoff}}<\infty$ plays the role of a formal cutoff for the IR region.\\
At the zero temperature, in the large $l_x$ limit, $r_t$ is expected to approach the location of the horizon $r_0$.  In all the three different phase, we have $g^{rr}g_{xx}\sim r^4$, ~
$g_{xx}g_{yy}g_{zz}\sim r^6$ in the UV, while $g^{rr}g_{xx}\sim r^{2+\alpha}$, ~$g_{xx}g_{yy}g_{zz}\sim r^{2\alpha+\delta}$ in the deep IR, with $2\alpha+\delta <6$. As a consequence, we have $
\mathfrak{C}_1=f_0\sqrt{u_0}\,r_t^{\alpha+\frac{\delta}{2}} $. We can numerically prove that the $l_x$ integral in \eqref{eq:ellx} is dominated by the IR region, i.e., the first term in \eqref{eq:ellx}.
Using Eqs. (\ref{eq:IRNLSM}), (\ref{eq:IRQCP}) and (\ref{eq:IRTrivial}), 
we find that
\be
l_x\simeq \frac{2}{\sqrt{f_0}}\int_{r_t}^{r_{\text{cutoff}}} \frac{dr}{r} \sqrt{\frac{r^{-\alpha}}{(\frac{r}{r_t})^{2\alpha+\delta}-1}}\,.
\ee
From the above equation, we have $l_x\propto r_t^{-\alpha/2}$ and therefore $r_t\propto l_x^{-2/\alpha}$. Exploiting the fact that $
\mathfrak{C}_1=f_0\sqrt{u_0}\,r_t^{\alpha+\frac{\delta}{2}} $, we derive
$\mathfrak{C}_1\propto l_x^{-2-{\bf z}}$ and $c_x\propto l_x^{1-{\bf z}}$.

Similarly, in the large $l_z$ limit, the dominant contribution in the integral of $l_z$ comes from the IR regime, expressed as
\bea
l_z\simeq \frac{2}{\sqrt{u_0}}\int_{r_t}^{r_{\text{cutoff}}}
\frac{dr}{r}\sqrt{\frac{r^{-\delta}}{(\frac{r}{r_t})^{2\alpha+\delta}-1}}\,.
\eea
The above equation leads to $l_z\propto r_t^{-\delta/2}$ and $r_t\propto l_z^{-2/\delta}$. 
Therefore, $\mathfrak{C}_3\propto l_z^{-1-2/{\bf z}}$ and $c_z\propto l_z^{2-2/{\bf z}}$.

\begin{figure}[h!]
  \centering
\includegraphics[width=0.47\textwidth]{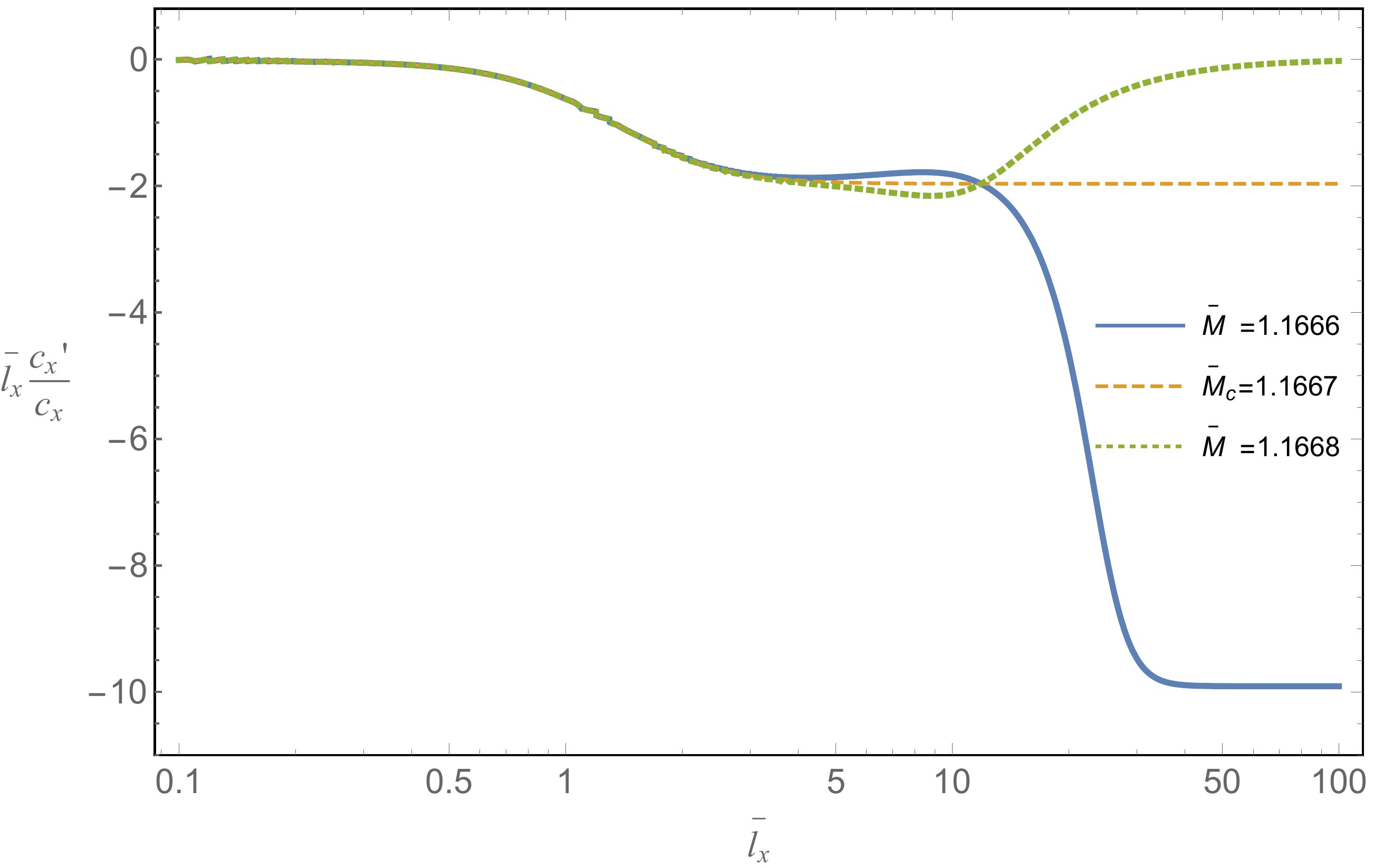}
~~~~~
\includegraphics[width=0.47\textwidth]{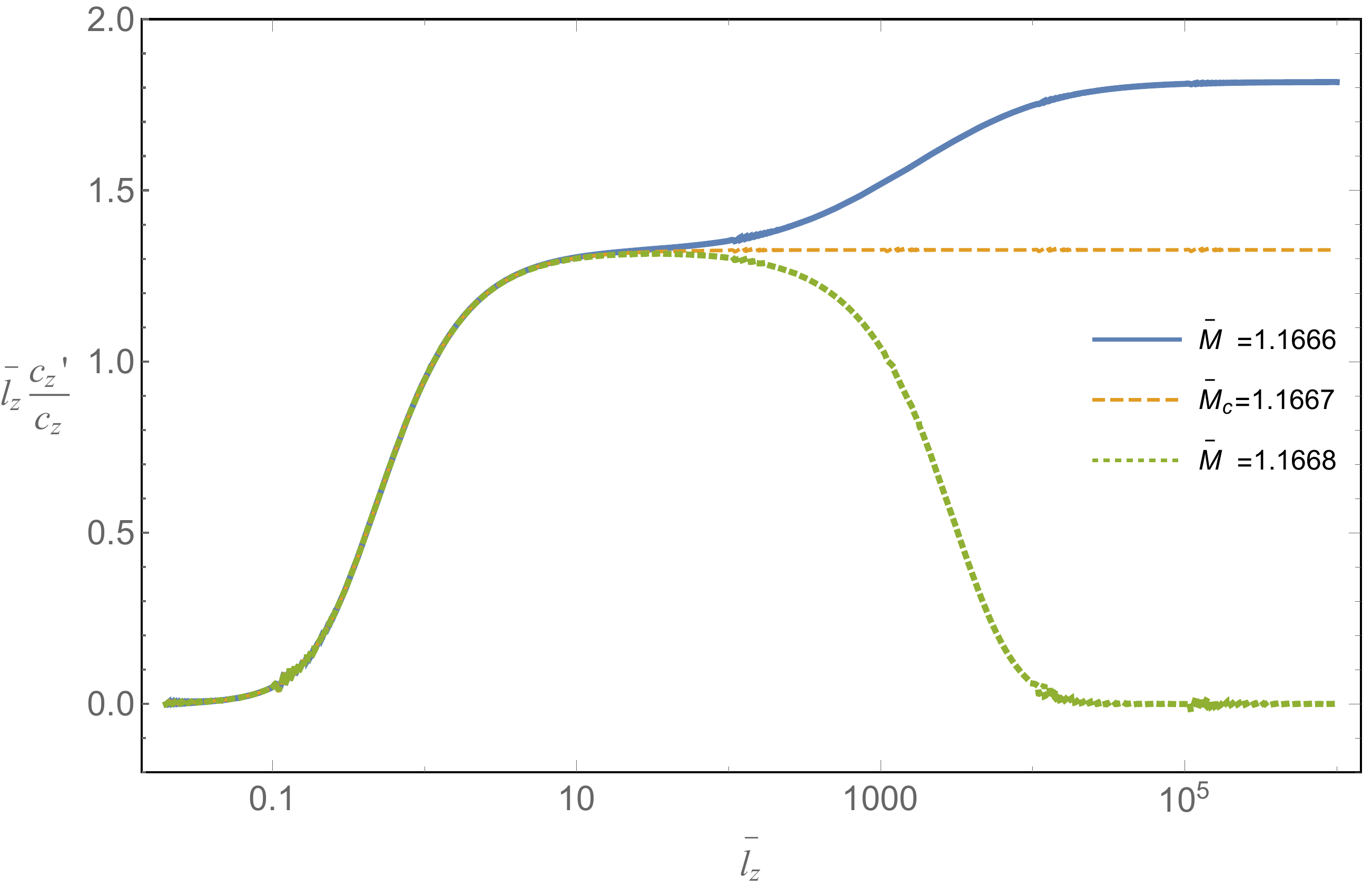}
  \caption{\small 
  Scaling properties of the c-functions $c_x$ ({\bf left}) 
  and $c_z$ ({\bf right}) 
  with respect to 
  $\bar{l}_x$ or $\bar{l}_z$. The different lines correspond to the three different IR phases, i.e., $\bar{M}= 1.1666$ (blue), $1.1667$ (orange dashed), $1.1668$ (green dotted). 
  }
  \label{fig:powerz}
\end{figure}

We have numerically confirmed the above results. These analytical scalings are summarized in Table \ref{table:lx} and Table \ref{table:lz} in the main text. One immediate observation is that one could use the $c$-functions to probe the scaling exponents related to the different IR phases. The nature of the UV and IR fixed points, and indeed the whole RG flow structure of the dual field theory, are reflected in the power law behaviors of the turning point, the entanglement entropy and the $c$ functions. Furthermore, as shown in Fig.\ref{fig:powerz}, the above analysis is also able to probe the intermediate geometry and the full RG flow trajectories. The scaling of the $c$-functions for small $\bar{l}_i$ is universal since it is given by the UV fixed point. On the contrary, the scaling for large $\bar{l}_i$ is a probe of the IR fixed point and depends crucially on $\bar{M}$. In the quantum critical state $\bar{M}_c=1.1667$, the scaling reflects the anisotropic Lifshitz symmetry with ${\bf z}=2.968$. Near the critical point, i.e. $\bar{M}=1.1666$ (NLSM phase) and $\bar{M}=1.1668$ (trivial phase), before reaching the large $l$ regime, the scaling exponents are those of the QCP and finally split into the NLSM (${\bf z}=10.908$) and trivial (${\bf z}=1$) branches. This behavior is qualitatively very similar to those of the bulk field along the radial direction shown in Fig.\ref{fig:profiles}. This is not a coincidence, since both quantities probe the RG flow structure of the theory where $r\rightarrow \infty,l_i\rightarrow 0$ represent the UV regime and $r\rightarrow 0,l_i \rightarrow \infty$ the IR.

\vspace{0.5cm}

\subsection{Analytical results and the relation between entanglement and the nodal line length}\label{app:B2}
Based on the discussion of the scaling behaviors in the previous subsection, we further show that in the NLSM phase there exists an analytic relation between $c_x$ ($c_z$) and $\mathcal{B}_{tz}$ ($\mathcal{B}_{xy}$).

From \eqref{eq:c-functions}, we have 
\be
c_x=\frac{1}{2}\mathfrak{C}_1l_x^3
=\frac{\sqrt{u(r_t)}f(r_t)}{2}l_x^3\,,~~~~
c_z=\frac{1}{2}\mathfrak{C}_3l_z^3=\frac{\sqrt{u(r_t)}f(r_t)}{2}l_z^3\,.
\ee
Meanwhile, 
\bea
\begin{split}
l_x\sqrt{f(r_t)}\simeq
2\int_{r_t}^{r_{\text{cutoff}}}
\frac{dr}{r}\sqrt{\frac{(\frac{r}{r_t})^{-\alpha}}{(\frac{r}{r_t})^{2\alpha+\delta}-1}}\,=2\int_{1}^{\tilde{r}_{\text{cutoff}}}
\frac{d\tilde{r}}{\tilde{r}}
\sqrt{\frac{\tilde{r}^{-\alpha}}{\tilde{r}^{2\alpha+\delta}-1}}\,=\frac{4\sqrt{\pi}}{\alpha}\frac{\Gamma(\frac{3\alpha+\delta}{4\alpha+2\delta})}{\Gamma(\frac{\alpha}{4\alpha+2\delta})}
\end{split}
\eea
and
\bea
\begin{split}
l_z\sqrt{u(r_t)}&\simeq
2\int_{r_t}^{r_{\text{cutoff}}}\frac{dr}{r}\sqrt{\frac{(\frac{r}{r_t})^{-\delta}}{(\frac{r}{r_t})^{2\alpha+\delta}-1}}\,
=\frac{4\sqrt{\pi}}{\delta}\frac{\Gamma(\frac{\alpha+\delta}{2\alpha+\delta})}{\Gamma(\frac{\delta}{4\alpha+2\delta})}\,,
\end{split}
\eea
where $\tilde{r}\equiv r/r_t$.

At the same time, we have
\be
\frac{\mathcal{B}_{tz}}{u}\Big{|}_{r_t}
=b_{tz0}=\sqrt{\frac{\alpha-\sqrt{\alpha^3\delta}}{2\lambda_2(\alpha-\delta)}}\,,~~~~
\frac{\mathcal{B}_{xy}}{f}\Big{|}_{r_t}=b_{xy0}=\sqrt{\frac{\delta-\sqrt{\alpha\delta^3}}{2\lambda_2(\alpha-\delta)}}\,.
\ee

Therefore, putting the above results together we finally obtain
\be
\frac{l_x^{-1}c_x}{\sqrt{\mathcal{B}_{tz}}|_{r_t}}
=\frac{16\pi}{\alpha^2}
\frac{\Gamma^2(\frac{3\alpha+\delta}{4\alpha+2\delta})}{\Gamma^2(\frac{\alpha}{4\alpha+2\delta})}\left(\frac{2\lambda_2(\alpha-\delta)}{\alpha-\sqrt{\alpha^3\delta}}\right)^{\frac{1}{4}}\,,~~~~~~~
\label{eq:ratio}
\frac{l_z^{-2}c_z}{\mathcal{B}_{xy}|_{r_t}}=
\frac{2\sqrt{\pi}}{\delta}
\frac{\Gamma(\frac{\alpha+\delta}{2\alpha+\delta})}{\Gamma(\frac{\delta}{4\alpha+2\delta})}
\sqrt{\frac{2\lambda_2(\alpha-\delta)}{\delta-\sqrt{\alpha\delta^3}}}\,.
\ee

\begin{figure}[h!]
  \centering
\includegraphics[width=0.45\textwidth]{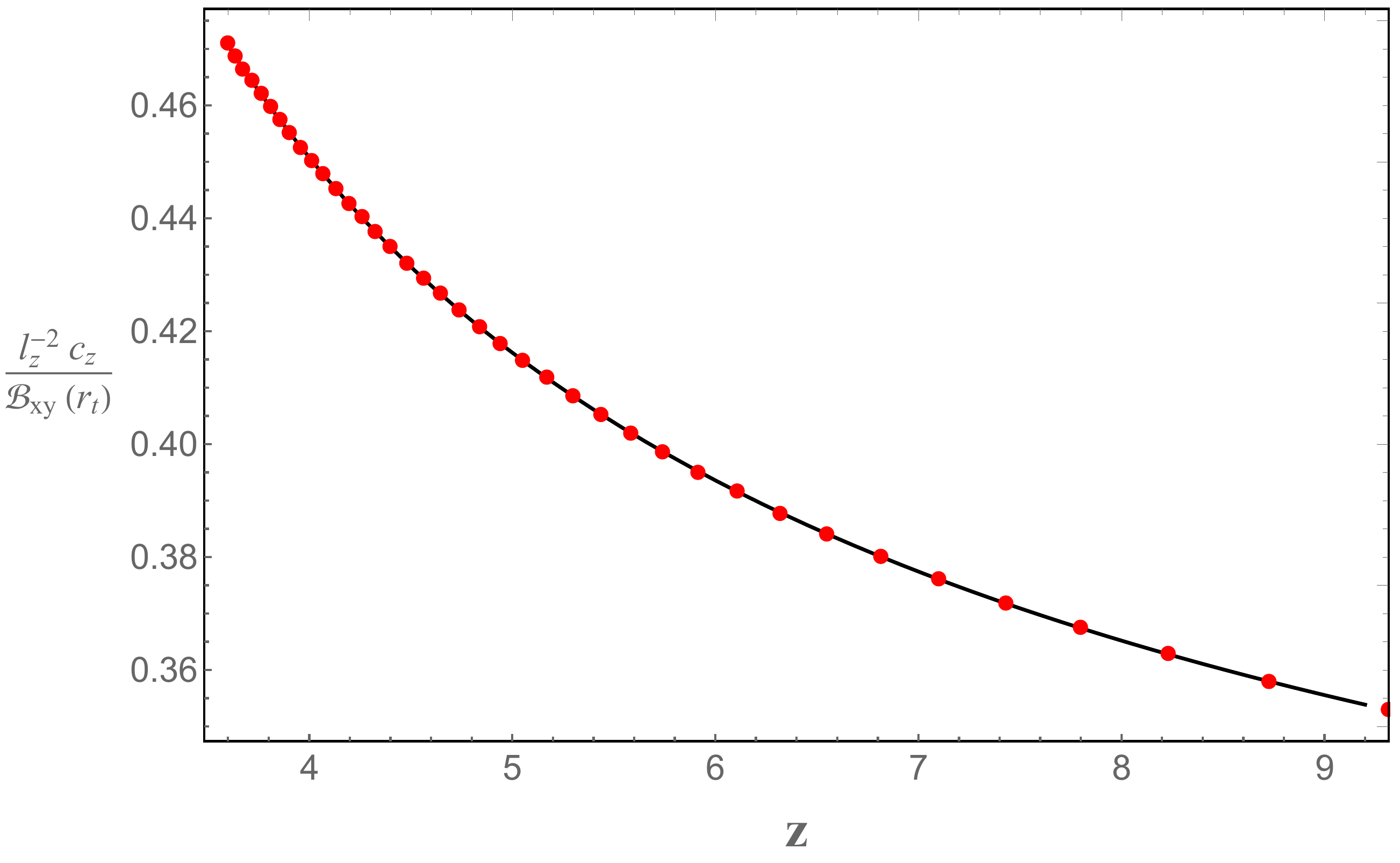}
\hspace{0.1pt}
\includegraphics[width=0.45\textwidth]{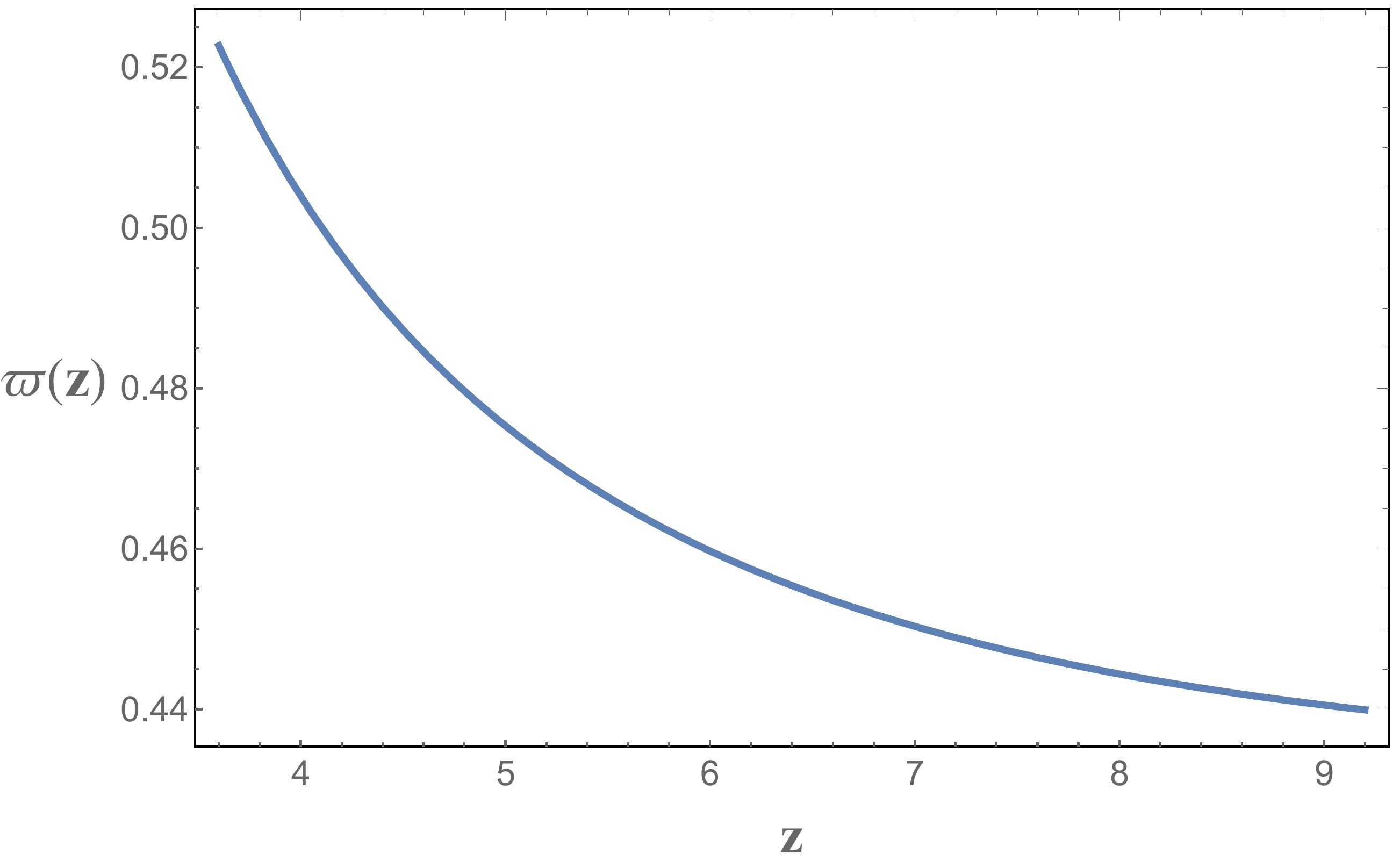}
  \caption{\small \textbf{Left: } The confirmation of the analytical expressions in \eqref{eq:ratio}. The numerical data (red dots) computed for $\bar{M}=0$ and different $\lambda_2$ coincide with the  analytical expression (black line) in \eqref{eq:ratio}. \textbf{Right:} $\varpi({\bf z})$ as a function of ${\bf z}$. }
  \label{fig:ratio}
\end{figure}

In the main text, we have defined $\varpi({\bf z})$ in \eqref{deforder}.  
From \eqref{eq:ratio}, we have 
\be
\varpi^{-1}({\bf z})=b_{xy0}\tilde{f}_0\left(\frac{16 \pi}{\tilde{u}_0\delta^2}\right)^{\frac{1}{{\bf z}}}
\left(\frac{\Gamma(\frac{\alpha+\delta}{2\alpha+\delta})}{\Gamma(\frac{\delta}{4\alpha+2\delta})}\right)^{\frac{2}{\bf z}}
\ee
in the nodal line semimetal phase, which indicates $\varpi({\bf z})$ remains a ${\bf z}$-dependent constant in the NLSM phase.  Here $\tilde{f}_0, \tilde{u}_0$ are values of $f_0, u_0$ at $\bar{M}=0$ that come from normalization in \eqref{eq:orderM} in the man text, and they only depend on ${\bf z}$. In the inset of Fig.\ref{fig:order1} in the main text, we numerically show that for a specific ${\bf z}=10.908$, $\varpi({\bf z}=10.908)$ is a $\bar{M}$-independent constant in the NLSM phase. The behavior of $\varpi({\bf z})$ as a function of ${\bf z}$ is shown in the right panel of Fig.\ref{fig:ratio}. 

At the quantum critical point, a similar analysis is also applicable and we only need to change the parameters to be defined at the critical point. 
However, in the topologically trivial phase, the leading order solutions for $B_{tz}$ and $B_{xy}$ vanish and no relations of such a form can be derived. The validity of these analytic expressions is verified using numerical data. For example, in Fig.\ref{fig:ratio} we show the agreement between \eqref{eq:ratio} and the numerical data. 

\bibliographystyle{JHEP}
\bibliography{refs}

\end{document}